\documentclass[11pt,a4paper]{article}

\usepackage{jheppub}
\usepackage{bm}

\def\GeV{\ifmmode {\mathrm{\ Ge\kern -0.1em V}}\else
                   \textrm{Ge\kern -0.1em V}\fi}%
\def\TeV{\ifmmode {\mathrm{\ Te\kern -0.1em V}}\else
                   \textrm{Te\kern -0.1em V}\fi}%

\let\gev=\GeV

\def\met{\ensuremath{E_{\mathrm{T}}^{\mathrm{miss}}}} 
\def\ttbar{\antibar{t}}
\def\antibar#1{\ensuremath{#1\bar{#1}}}
\def\pt{\ensuremath{p_{\mathrm{T}}}} 
\def\et{\ensuremath{E_{\mathrm{T}}}} 
\def\HT{\ensuremath{H_{\mathrm{T}}}} 
\def\MET{\ensuremath{E_{\mathrm{T}}^{\mathrm{miss}}}} 
\def\met{\ensuremath{E_{\mathrm{T}}^{\mathrm{miss}}}} 
\def\metx{\ensuremath{E_{x}^{\mathrm{miss}}}} 
\def\mety{\ensuremath{E_{y}^{\mathrm{miss}}}} 

\newcommand{\RE}{\text{Re}\,}

\newcommand{\fz}{F_{\rm 0}}
\newcommand{\fl}{F_{\rm L}}
\newcommand{\fr}{F_{\rm R}}
\newcommand{\costs}{\cos \theta^{*}}

\newcommand{\ap}{A_{\rm +}}
\newcommand{\am}{A_{\rm -}}
\newcommand{\apm}{A_{\rm \pm}}

\newcommand{\vl}{V_{\rm L}}
\newcommand{\vr}{V_{\rm R}}
\newcommand{\gl}{g_{\rm L}}
\newcommand{\gr}{g_{\rm R}}

\newcommand{\AKT}{anti-$k_t$}

\usepackage{preprintcover}  
\PreprintCoverPaperTitle{Measurement of the $\bm{W}$ boson polarization in top quark decays with the
ATLAS detector}  
\PreprintIdNumber{CERN-PH-EP-2012-112}  
\PreprintCoverAbstract{This paper presents measurements of the polarization of $W$
  bosons in top quark decays, derived from \ttbar\ events with missing
  transverse momentum, one charged lepton and at least four jets, or
  two charged leptons and at least two jets. Data from $pp$ collisions
  at a centre-of-mass energy of 7~\TeV\ were collected with the ATLAS
  experiment at the LHC and correspond to an integrated luminosity of
  1.04~fb$^{-1}$.  The measured fractions of longitudinally, left- and
  right-handed polarization are $\fz = 0.67 \pm 0.07$, $\fl = 0.32 \pm
  0.04$ and $\fr = 0.01 \pm 0.05$, in agreement with the Standard Model
  predictions. As the polarization of the $W$ bosons in top quark
  decays is sensitive to the $Wtb$ vertex Lorentz structure and couplings, the
  measurements were used to set limits on anomalous contributions to
  the $Wtb$ couplings.}  
\PreprintJournalName{JHEP}  

\title{Measurement of the $\bm{W}$ boson polarization in top quark decays with the
ATLAS detector}

\author[a]{The ATLAS Collaboration}
\affiliation[a]{CERN, Geneva}
\emailAdd{atlas.publications@cern.ch}

\abstract{This paper presents measurements of the polarization of $W$
  bosons in top quark decays, derived from \ttbar\ events with missing
  transverse momentum, one charged lepton and at least four jets, or
  two charged leptons and at least two jets. Data from $pp$ collisions
  at a centre-of-mass energy of 7~\TeV\ were collected with the ATLAS
  experiment at the LHC and correspond to an integrated luminosity of
  1.04~fb$^{-1}$.  The measured fractions of longitudinally, left- and
  right-handed polarization are $\fz = 0.67 \pm 0.07$, $\fl = 0.32 \pm
  0.04$ and $\fr = 0.01 \pm 0.05$, in agreement with the Standard Model
  predictions. As the polarization of the $W$ bosons in top quark
  decays is sensitive to the $Wtb$ vertex Lorentz structure and couplings, the
  measurements were used to set limits on anomalous contributions to
  the $Wtb$ couplings.}


\begin{document}
\maketitle


\section{Introduction}
\label{sec:introduction}

Discovered in 1995 by the CDF and D\O\ experiments
\cite{Abe:1995hr,Abachi:1995iq}, the top quark is the heaviest known
fundamental particle, with a mass of
$173.2\pm0.9$~\GeV~\cite{TevatronElectroweakWorkingGroup:2011wr}.
Measurements of top quark properties play an important role in testing
the Standard Model and its possible extensions. One particular test is
the study of the $Wtb$ vertex Lorentz structure and couplings, 
which can be probed by measuring the polarization of $W$ bosons 
produced in top quark decays.

At the LHC, top quarks are produced mainly in pairs via the strong
interaction and are predicted to decay via the electroweak interaction
into a $W$ boson and a bottom quark with a nearly 100\% branching
fraction. Events with $t\bar{t}$ pairs can thus be classified
according to the decay of the two produced $W$ bosons. Each boson can
decay either into a quark-antiquark pair or into a charged lepton and
a neutrino. The single-lepton and dilepton topologies, both considered
in the analyses presented in this paper, have one and two isolated
charged leptons in the final state. Only electrons and muons,
including those from $\tau$ decays, are considered here.

The $Wtb$ vertex is defined by the electroweak interaction and has a
$(V-A)$ structure where $V$ and $A$ are the vector and axial-vector
contributions to the vertex. Since the $W$ bosons are produced as real
particles in top quark decays, their polarization can be longitudinal,
left-handed or right-handed. The fractions of events with a particular
polarization, $\fz$, $\fl$ and $\fr$, are referred to as helicity
fractions. They are predicted in next-to-next-to-leading-order (NNLO)
QCD calculations to be $\fz = 0.687\pm0.005$, $\fl = 0.311\pm0.005$,
$\fr = 0.0017\pm0.0001$~\cite{Czarnecki:2010gb}. These fractions can
be extracted from measurements of the angular distribution of the
decay products of the top quark. The angle $\theta^{*}$ is defined as
the angle between the momentum direction of the charged lepton from
the decay of the $W$ boson and the reversed momentum direction of the
$b$-quark from the decay of the top quark, both boosted into the $W$
boson rest frame~\cite{Kane:1991bg}. The angular distribution is:
\begin{equation}
\frac{1}{\sigma} \frac{{\rm d} \sigma}{{\rm d} \costs} = \frac{3}{4} \left( 1-\cos^{2} \theta^{*} \right) \, \fz + \frac{3}{8} \left(1-\costs \right)^2 \, \fl + 
\frac{3}{8} 
\left(1 + 
\costs\right)^{2} \, \fr \,.
\label{ec:dist}
\end{equation}
All previous measurements of the helicity fractions, performed by the
CDF and
D\O\ Collaborations~\cite{Aaltonen:2012rz,Aaltonen2010105,Abazov:2010jn}
at the Tevatron, are in agreement with Standard Model predictions.

Information about the polarization of the $W$ bosons can also be
obtained through complementary observables, such as the angular
asymmetries, $\ap$ and $\am$, defined as:
\begin{equation}
A_{\pm} = \frac{N(\costs > z) - N(\costs < z)}{N(\costs > z) +
N(\costs < z)} \,,
\label{eq:asym}
\end{equation}
with $z=\pm(1-2^{2/3})$ for $\apm$, allowing the dependence on $\fl$
and $\fr$ to cancel, respectively.  The asymmetries can be related to
the helicity fractions by a simple system of
equations~\cite{delAguila:2002nf,AguilarSaavedra:2006fy}. In the
Standard Model, the NNLO values for these asymmetries are $\ap =0.537
\pm 0.004$ and $\am =-0.841 \pm 0.006$~\cite{Czarnecki:2010gb}.

In the presence of anomalous $Wtb$ couplings the helicity fractions
and angular asymmetries depart from their Standard Model 
values~\cite{Kane:1991bg,AguilarSaavedra:2006fy}. In
effective field theories, dimension-six operators can be introduced
which modify the
$Wtb$ vertex~\cite{Buchmuller:1985jz,AguilarSaavedra:2008zc,Zhang:2010dr}.
Coefficients controlling the strength of these operators can be
constrained by measurements of the helicity fractions or the angular
asymmetries.

This paper describes measurements of the $W$ boson polarization in top
quark decays and the constraints on the $Wtb$ vertex structure based
on a data set recorded with the ATLAS detector between March and June
2011 and corresponding to an integrated luminosity of
1.04~fb$^{-1}$. The helicity fractions were measured using two
different methods. The first compares the observed $\costs$
distribution with templates for different $W$ boson helicity states
obtained from simulation. The second method extracts angular
asymmetries from an unfolded $\costs$ spectrum corrected for
background contributions. Limits on anomalous couplings, generated by
the aforementioned dimension-six operators, were set using the
combined result from the two measurements.

This paper is structured as follows. Data samples and simulated
samples modelling signal and background processes are described in
Section~\ref{sec:data}, followed by a summary of the event selection
in Section~\ref{sec:selection}. The analysis strategies can be found
in Section~\ref{sec:measurements}. Sources of systematic uncertainty
are discussed in Section~\ref{sec:systematics}, the results are
summarized in Section~\ref{sec:results} and
Section~\ref{sec:conclusions} concludes the paper.


\section{Data and simulation samples}
\label{sec:data}

The ATLAS detector~\cite{atlaspaper} at the LHC covers nearly the entire 
solid angle around the collision point. It consists of an inner tracking 
detector surrounded by a thin superconducting solenoid, electromagnetic 
and hadronic calorimeters, and an external muon spectrometer 
incorporating three large superconducting toroid magnet assemblies. A 
three-level trigger system, designed to reduce the event rate from 
40~MHz to about 200\,Hz, is used to select events of interest.

\subsection{Data sample}

Data from $pp$ collisions at a centre-of-mass energy of 7~\TeV\ were
collected using single-lepton triggers with transverse momentum
thresholds of 20~\GeV\ for electrons and 18~\GeV\ for muons. The size
of the data sample corresponds to an integrated luminosity of
$1.04\pm0.04$~fb$^{-1}$~\cite{lumi,Aad:2011dr}.

\subsection{Signal and background modelling}

Signal and most background processes were modelled by Monte Carlo (MC)
simulation. The signal process was simulated using two different
generators. The leading order (LO) Monte Carlo generator {\sc
  Protos}~\cite{AguilarSaavedra:2006fy,AguilarSaavedra:2008gt} was used with
the CTEQ6L1 sets of parton distribution functions (PDFs)~\cite{cteq6l}
to generate three samples assuming $\fz=1$, $\fl=1$ and
$\fr=1$. This was achieved by choosing appropriate values for
anomalous $Wtb$ couplings, as described in
Refs.~\cite{delAguila:2002nf,AguilarSaavedra:2006fy}.  The output was
interfaced to {\sc Pythia}~\cite{Sjostrand:2006za} to simulate parton
showers and hadronization. In addition, the next-to-leading-order
(NLO) generator {\sc
  MC@NLO}~\cite{mcatnlo,Frixione:2003ei,Frixione:2005vw} was used with
the CTEQ6.6 PDF set~\cite{CTEQ66} for studies of systematic
differences in the top quark production and decay modelling. 

The $W$ and $Z$ boson production in association with multiple jets was
simulated using the {\sc Alpgen} generator~\cite{Alpgen} and the
CTEQ6L1 PDF set. Tree-level matrix elements with up to five final-state 
partons were included. The standard {\sc Alpgen} matching scheme
was used to remove overlaps between the $n$ and $n+1$ parton
samples. Heavy flavour samples containing $Wb\bar{b}$, $Wc\bar{c}$,
$Wc$ and $Zb\bar{b}$ events were simulated separately. The $Z$+jets
samples were generated with dileptons in the invariant mass range $10
< m_{\ell\ell} < 2000$~\gev.  Diboson processes were simulated using
{\sc Herwig}~\cite{herwig}. Single top quark production was simulated
using {\sc MC@NLO}, invoking the `diagram removal
scheme'~\cite{Frixione:2008yi} to remove overlaps between the single
top quark and $\ttbar$ final states. An additional sample of $W$+jets
events was generated with {\sc Sherpa}~\cite{Gleisberg:2008ta} and
used to study systematic uncertainties. Details are described in
Section~\ref{sec:systematics}.
Apart from the {\sc Protos} \ttbar\ samples, all events were
hadronized with {\sc Herwig} using {\sc
Jimmy}~\cite{Butterworth:1996zw} for the underlying event model. Different
underlying event tunes were used, depending on the hadronisation program used
({\em i.e.} {\sc Pythia} or {\sc Herwig})~\cite{PUB-2011-008}.

Additional MC samples were used for the evaluation of systematic
uncertainties. These were generated with the {\sc
  AcerMC}~\cite{AcerMC35} and {\sc
  Powheg}~\cite{Nason:2004rx,Frixione:2007vw} generators interfaced to
{\sc Pythia} for hadronization. They are described in
Section~\ref{sec:systematics}. A top quark mass of $172.5$~\gev\ was
assumed for all signal samples if not stated otherwise. Additional
{\sc MC@NLO} samples were generated assuming different top quark
masses.

All simulated events contain multiple $pp$-interactions and contain 
pileup contributions corresponding to a bunch spacing of 50 ns. These 
simulated events were re-weighted such that the average number of 
interactions per proton-proton bunch crossing was the same in data and MC 
simulation. The average number of interactions per event was roughly six. 
All samples were processed by the detector and trigger simulation after 
event generation~\cite{geant4,atlsim}, and subjected to the same 
reconstruction algorithms as the data.

The cross-section of simulated \ttbar\ samples was normalized to 
164.6~pb, the value obtained from approximate NNLO 
calculations~\cite{Moch:2008qy,Langenfeld:2009tc,Aliev:2010zk}, and 
consistent with recent measurements~\cite{ATLAS:2012aa}. While the 
$Z$+jets normalization was obtained from the NNLO QCD cross-section 
calculations in the single lepton channels, a data-driven estimate for 
the normalization was used for the dilepton 
channels~\cite{ATLAS:2012aa}. The diboson with jets production was 
rescaled to match NLO calculations of the inclusive cross-sections.

For the single-lepton analysis, the multijet production background,
where an electron or muon originates from hadron decay or instrumental
background, was estimated from data as described in
Refs.~\cite{Collaboration:2010ey,chargeasymmetry}. The $W$+jets
background was obtained from simulation, except that its normalization
was derived from data, from a study of the asymmetry in the production
of $W^+$ and $W^-$ bosons~\cite{chargeasymmetry}. For the dilepton
analysis, background contributions from $W$+jets, single-lepton \ttbar\
and single top quark production were estimated using the method
described in Ref.~\cite{ATLAS:2012aa}. In the single-lepton and
dilepton analyses this source of background is labelled as
``misidentified leptons''.


\section{Event selection}
\label{sec:selection}

\subsection{Physics object definition}

The reconstruction and identification of electrons, muons, jets and the 
magnitude of the missing transverse momentum ($\met$) used in the 
analyses presented here followed the criteria employed for the 
measurement of the $t\bar t$ production 
cross-section~\cite{Collaboration:2010ey}. More details, including a 
description of the trigger and the vertex requirements, can be found in 
Ref.~\cite{chargeasymmetry}.

The $t\bar t$ events with at least one isolated charged lepton
(electron or muon) in the final state were considered as
signal. Events with $\tau$ leptons decaying into muons or electrons in
the final state were also considered to be part of the signal.

All the considered events were required to fulfil general event
quality criteria and to have a well-defined primary vertex with at
least five associated tracks. Reconstructed electrons were required to
have $\et>25~\gev$ and $|\eta|<2.47$ excluding $1.37<|\eta|<1.52$,
while muons were required to have $\pt>20~\gev{}$ and
$|\eta|<2.5$~\footnote{ATLAS uses a right-handed coordinate system
  with its origin at the nominal interaction point (IP) in the centre
  of the detector and the $z$-axis along the beam pipe. The $x$-axis
  points from the IP to the centre of the LHC ring, and the $y$-axis
  points upward. Cylindrical coordinates $(r,\phi)$ are used in the
  transverse plane, $\phi$ being the azimuthal angle around the beam
  pipe.  The pseudorapidity is defined in terms of the polar angle
  $\theta$ as $\eta=-\ln\tan(\theta/2)$.}. Jets were reconstructed
with the \AKT\ algorithm~\cite{Cacciari:2008gp} with a radius
parameter equal to $0.4$. They were required to have
$\pt>25$~\gev\ and $|\eta|<2.5$. For electrons, the energy not
associated to the electron cluster but contained in a cone of $\Delta
R = \sqrt{(\Delta\phi)^2 + (\Delta\eta)^2} = 0.2$ around it was
required not to exceed 3.5~\gev{}. For muons, the sum of track
transverse momenta and the total energy deposited in a cone of $\Delta
R =0.3$ around the muon were both required to be less than
4~\gev{}. Muons reconstructed within a $\Delta R =0.4$ cone of a jet
satisfying $\pt > 20~\gev$ were removed to reduce the contamination
caused by muons from hadron decays within jets. Subsequently, jets
within $\Delta R =0.2$ of an electron candidate were removed to avoid
double counting, which can occur because electron clusters are also
reconstructed as jets.

A combination of two algorithms was used for $b$-jet identification
($b$-tagging). They are based on reconstructed secondary vertices and
the impact parameter significances of tracks within jets. The chosen
working point resulted in a $b$-tagging efficiency of 70\% for jets
originating from $b$-quarks in a sample of simulated $\ttbar$ events
and a light-quark jet rejection factor of about
100~\cite{ATLAS-CONF-2011-102}.

\subsection{Single-lepton channels}

The single-lepton $\ttbar$ signal is characterized by a high-$\pt$ 
isolated charged lepton and missing transverse momentum from the 
neutrino, from the leptonically decaying $W$ boson, two light-quark jets 
from the hadronically decaying $W$ boson and two $b$-quark jets. The two 
channels with either an electron or a muon in the final state are 
referred to as single-electron and single-muon channels. The following 
event selection requirements were applied:

\begin{itemize}
\item the appropriate single-electron or single-muon trigger had
  fired;
  
\item events were required to contain exactly one isolated electron or
  muon;
  
\item in the single-electron channel, $\met>35$~\gev\ and
  $m{_\mathrm{T}}(W)>25$~\gev\ were required\footnote{Here
  $m_\mathrm{T}(W)$ is the $W$ boson transverse mass, reconstructed as
  $\sqrt{ 2 \pt^{\ell} \pt^{\nu} \left[ 1-\cos ( \phi^{\ell} -
      \phi^{\nu} ) \right] }$ where the measured missing transverse
  momentum provides the neutrino information.  } while in the
  single-muon channel the criteria were $\met>20$~\gev\ and
  $\met+m{_\mathrm{T}}(W)>60$~\gev;

\item events were required to have at least four jets, with at
least one of them being tagged as a $b$-jet.
\end{itemize}

\begin{figure}
  \begin{center}
    \begin{tabular}{cc}     
      \includegraphics[width=0.45\textwidth]{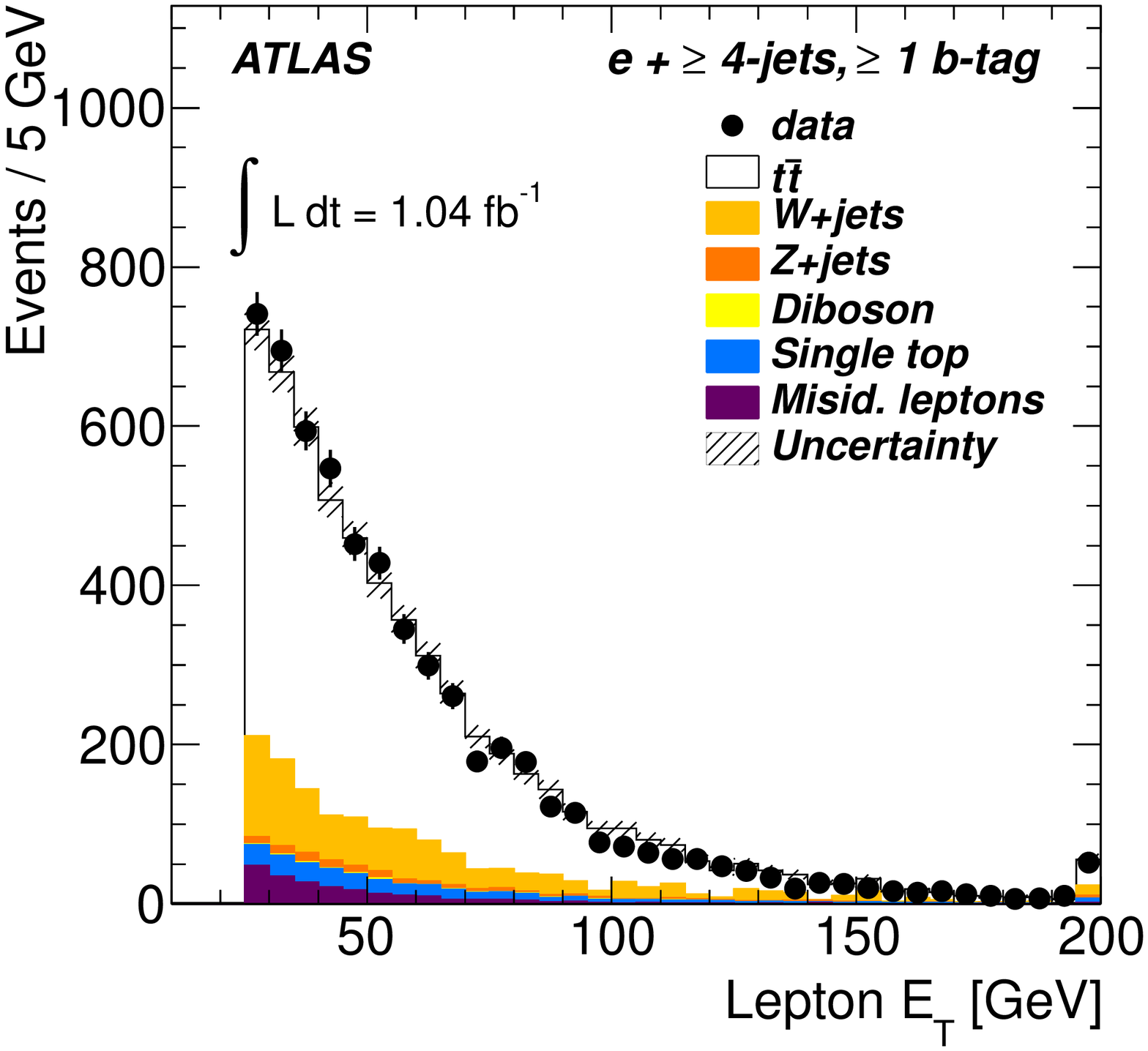} &
      \includegraphics[width=0.45\textwidth]{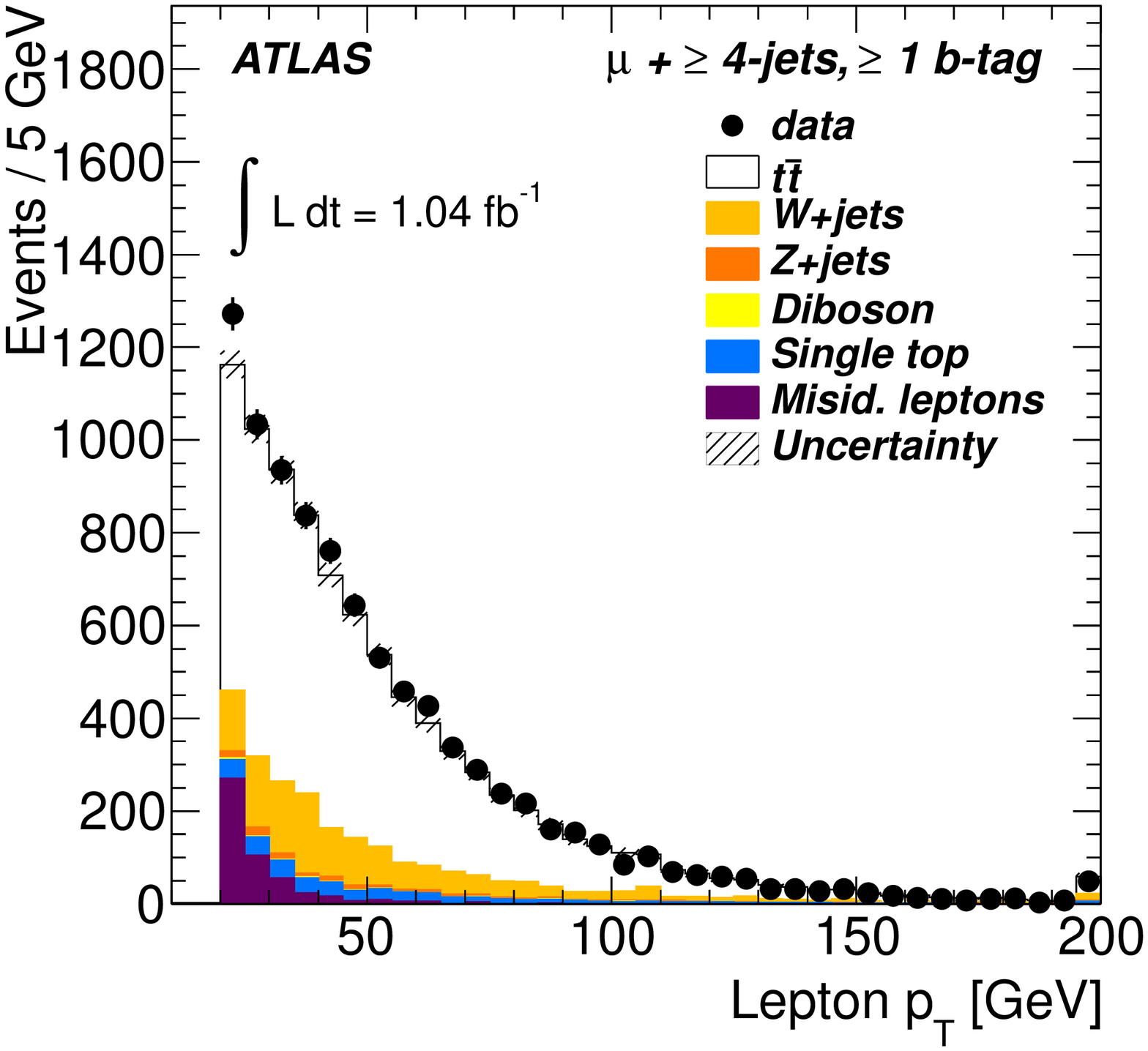} \\
      \includegraphics[width=0.45\textwidth]{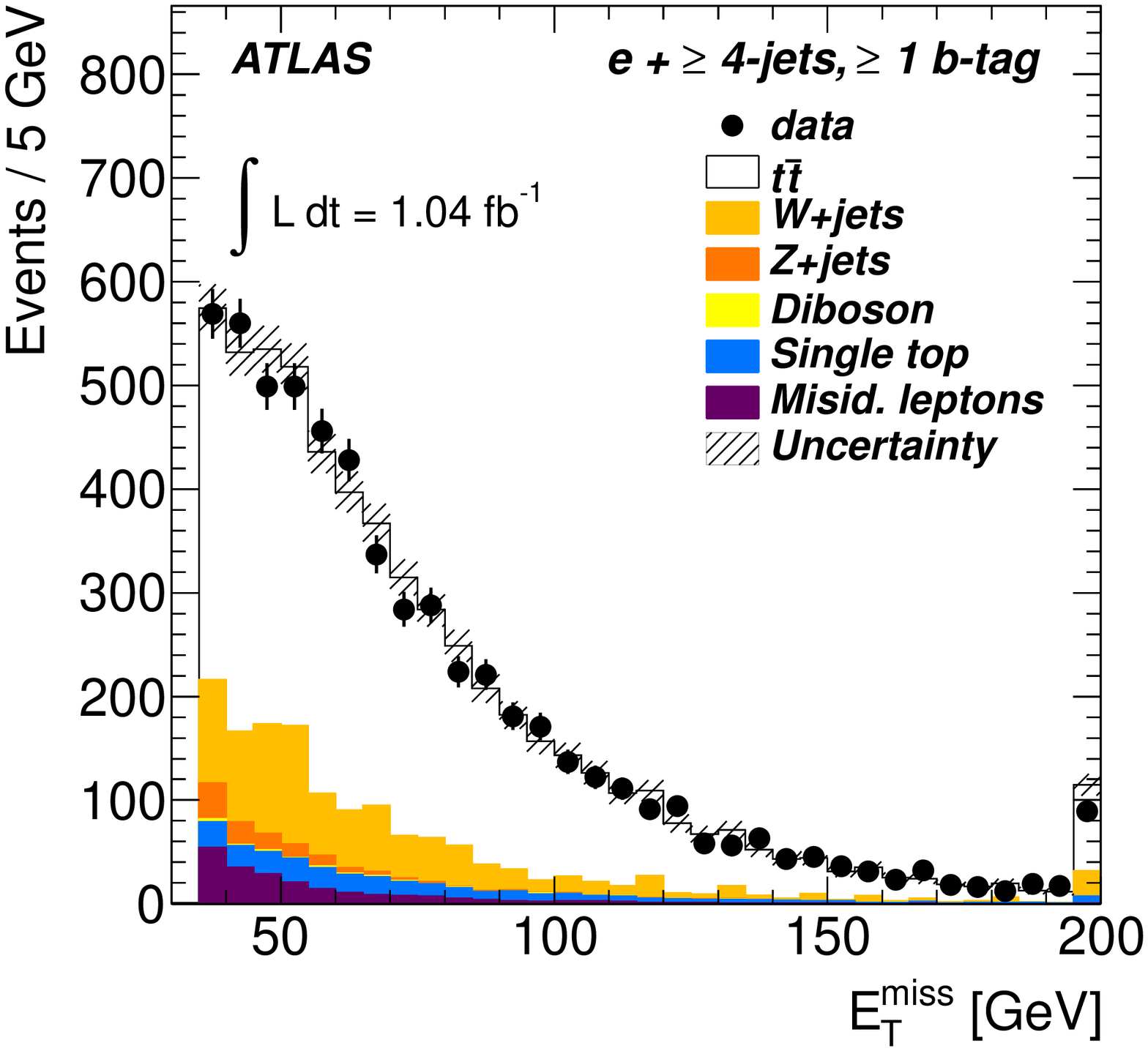} &
      \includegraphics[width=0.45\textwidth]{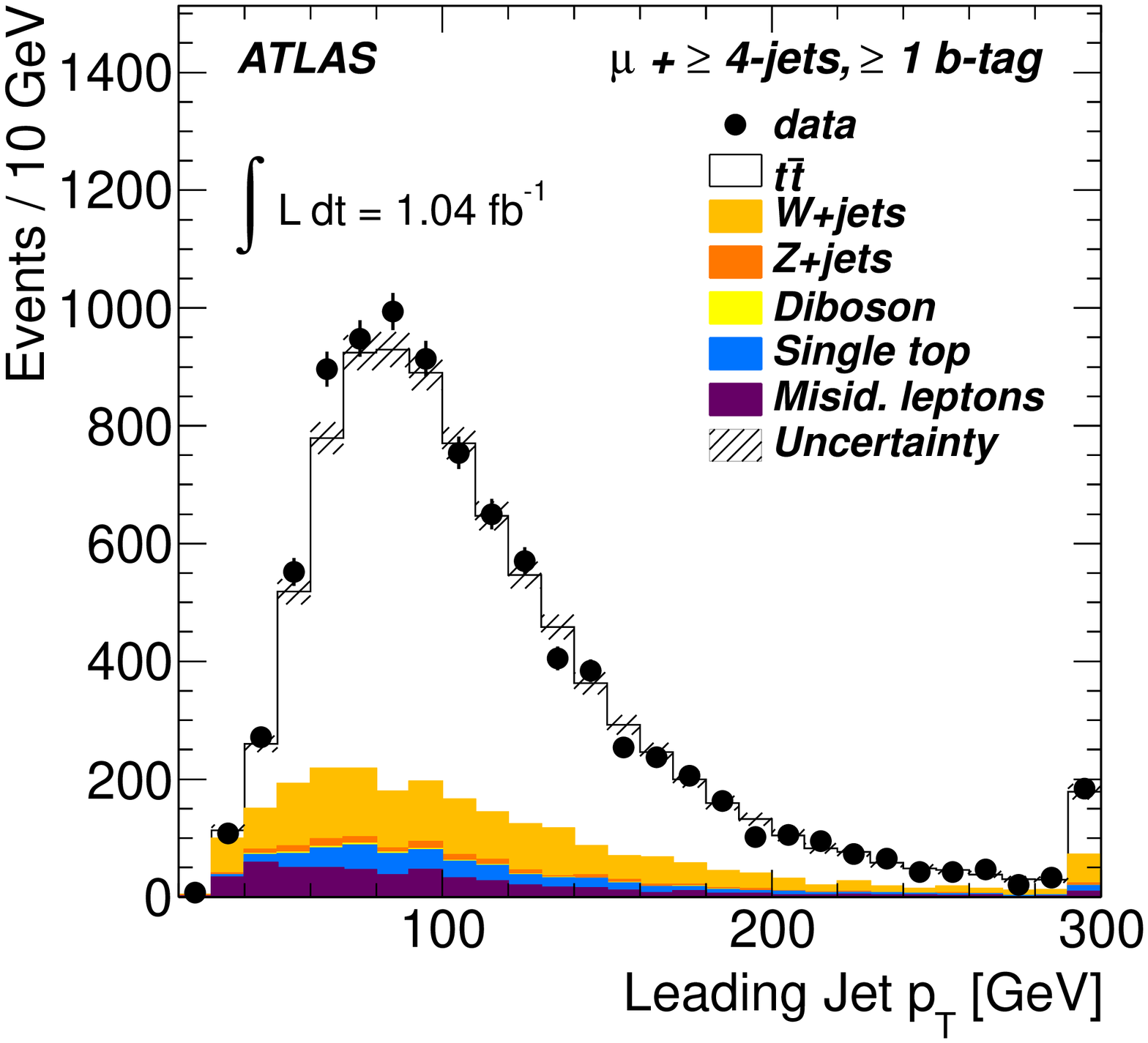}
    \end{tabular}
  \end{center}
  \caption{The top row shows (left) the $\et$ spectrum of the electron
    in the single-electron channel and (right) the $\pt$ spectrum of
    the muon in the single-muon channel. The bottom row shows (left)
    the missing transverse momentum in the single-electron channel and
    (right) the $\pt$ spectrum of the leading jet in the single-muon
    channel. The error band shows the statistical uncertainty from the
    Monte Carlo simulation. The last bin in each distribution includes
    the overflow.
    \label{fig:control_ljets}}
\end{figure}

The numbers of events expected after all selection requirements are
shown in Table~\ref{tab:yield_ljets} for the single-electron and
single-muon channels, together with the observed numbers of events in
the data sample. The single-electron channel contains significantly
fewer events than the single-muon channel due to the more stringent
requirements on the lepton \et\ (\pt), \met\ and $m{_\mathrm{T}}(W)$.
The observed numbers of events are in agreement with those expected,
with purities of 75\% and 72\% for the single electron and single muon
channels.  The systematic uncertainty discussed in
Section~\ref{sec:systematics} dominates the error on the event yields,
with jet energy scale, signal and background modelling, and
$b$-tagging being the dominant sources of uncertainty. Example
distributions in data are compared with the Standard Model predictions
in Figure~\ref{fig:control_ljets}. The data are described well by the
expected distributions.

\begin{table}
  \caption{Event yields in the single-electron and single-muon
    channels after the event selection. The
    table shows the expected number of events including their
    estimated total uncertainty as well as the number of events
    observed in the data sample.
    \label{tab:yield_ljets}} \begin{center}
    
    \begin{tabular}{llllll}
      \hline
	    Process & Single electron                                   & Single muon  \\
      \hline
      \ttbar\                 & $           4400 \pm           1100 $ & $           6500   \pm           1400 $ \\
      $W$+jets                & $ \phantom{0}900 \pm \phantom{0}700 $ & $           1400   \pm           1000 $ \\
      $Z$+jets                & $ \phantom{0}120 \pm \phantom{00}90 $ & $ \phantom{0}140   \pm \phantom{00}90 $ \\
      Diboson                 & $ \phantom{00}14 \pm \phantom{00}12 $ & $ \phantom{00}22   \pm \phantom{00}12 $ \\
      Single top              & $ \phantom{0}260 \pm \phantom{00}90 $ & $ \phantom{0}360   \pm \phantom{0}110 $ \\
      Misidentified leptons   & $ \phantom{0}220 \pm \phantom{0}220 $ & $ \phantom{0}500   \pm \phantom{0}500 $ \\
      \hline
      Total predicted & $ 5900 \pm 1300 $                     & $ 9000 \pm 1800 $ \\
      \hline
      Data            &  $5830$                               & $9121$     \\
      \hline
    \end{tabular}
  \end{center}
\end{table}

\subsection{Dilepton channels}

The dilepton $\ttbar$ signal is characterized by two high-$\pt$ isolated 
charged leptons, missing transverse momentum corresponding to the 
undetected neutrinos from the two leptonically decaying $W$ bosons, and 
two $b$-quark jets. The three channels are referred to as $ee$, $\mu\mu$ 
and $e\mu$ channels. The following event selection requirements were 
applied:

\begin{itemize}
\item the single-electron trigger had fired for the $ee$ channel, the 
single-muon trigger had fired for the $\mu\mu$ channel and
either of these triggers had fired for the $e\mu$ channel;

\item events were required to contain exactly two oppositely charged
  and isolated leptons ($ee$, $\mu\mu$ or $e\mu$);

\item at least two jets, with at least one of them being $b$-tagged,
  were required;

\item in order to avoid the low-mass Drell-Yan background region, events
  were required to have $m_{\ell\ell}>15\GeV$;

\item in the $ee$ and $\mu\mu$ channels, the missing transverse
  momentum had to satisfy $\MET>40\GeV$, and the invariant mass of the
  two leptons had to differ by at least $10\GeV$ from the $Z$ boson
  mass, {\em i.e.}\/ $|m_{\ell\ell}-m_Z|>10\GeV$, with $m_Z=91\GeV$,
  in order to suppress backgrounds from $Z$+jets events and events 
  containing misidentified leptons;

\item in the $e\mu$ channel, no \MET\ or $Z$ boson mass veto cuts were
  applied; however, the event \HT, defined as the scalar sum of the
  transverse energies of the two charged leptons and all selected
  jets, must satisfy $\HT>130\GeV$ to suppress backgrounds from
  $Z$+jets production;

 \item events were required to pass the $t\bar t$ reconstruction 
   criteria, as described in Section~\ref{sec:template}.

\end{itemize}

\begin{figure}
  \begin{center}
    \begin{tabular}{cc}     
      \includegraphics[width=0.45\textwidth]{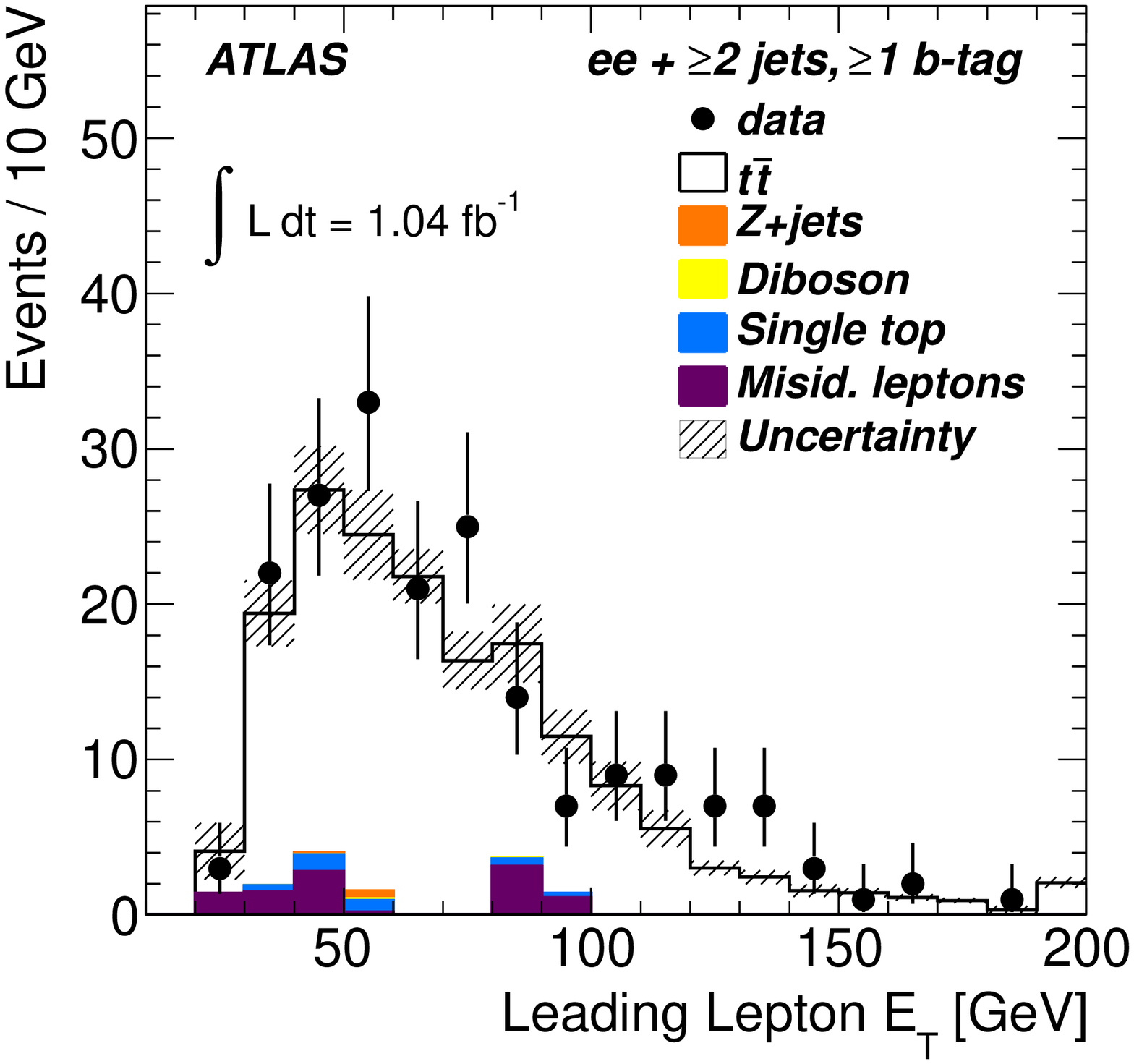} &
      \includegraphics[width=0.45\textwidth]{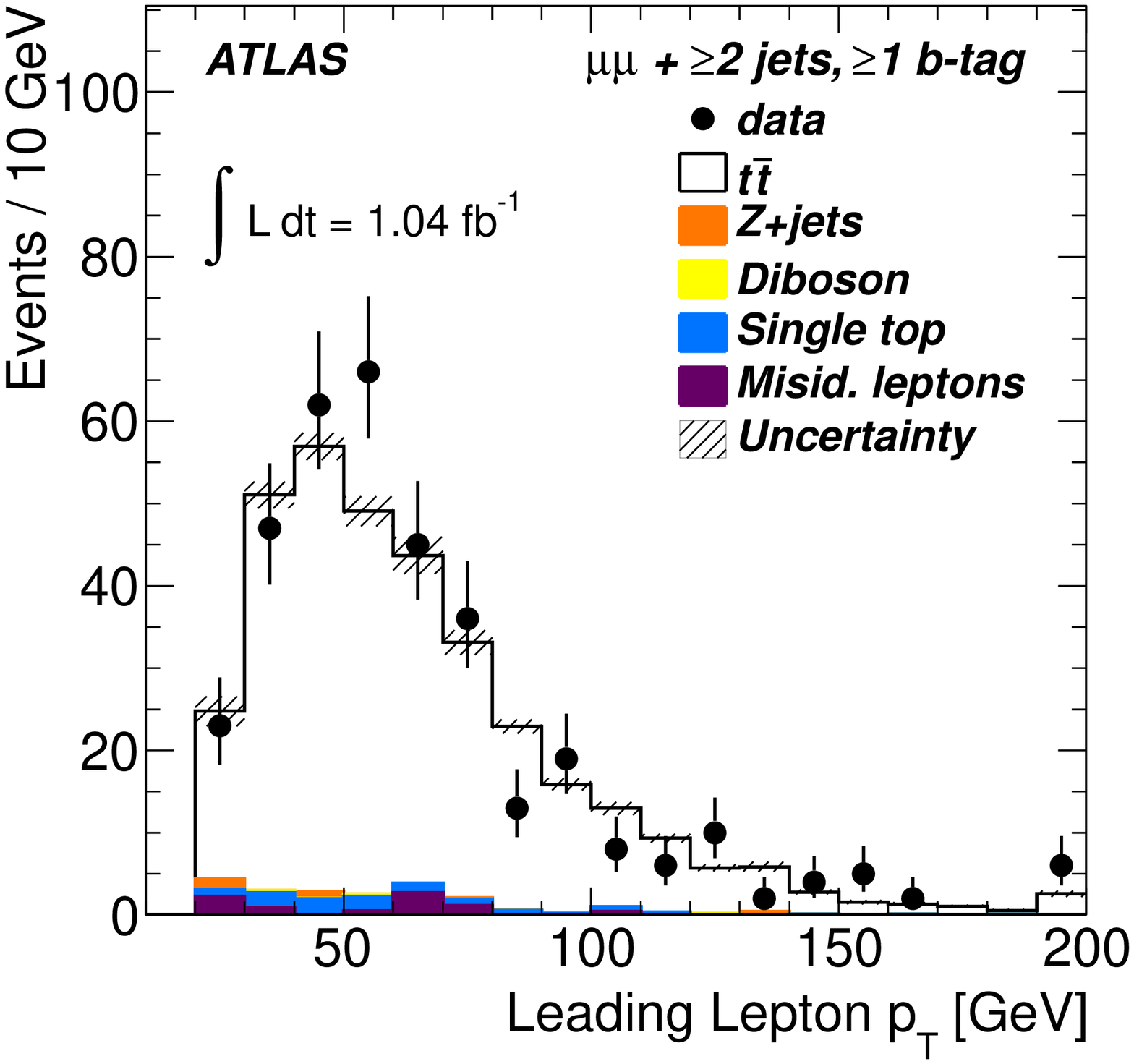} \\
      \includegraphics[width=0.45\textwidth]{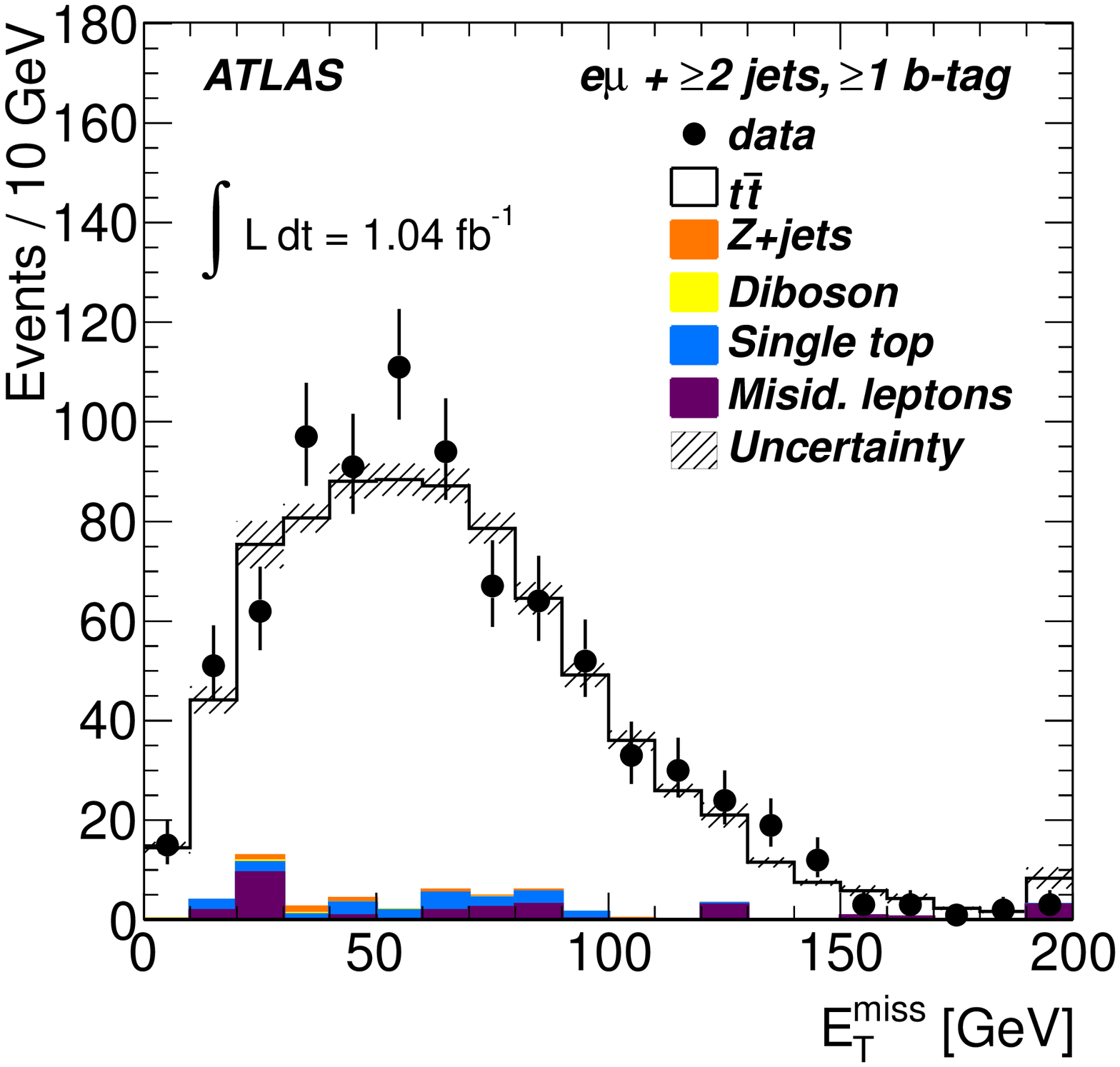} &
      \includegraphics[width=0.45\textwidth]{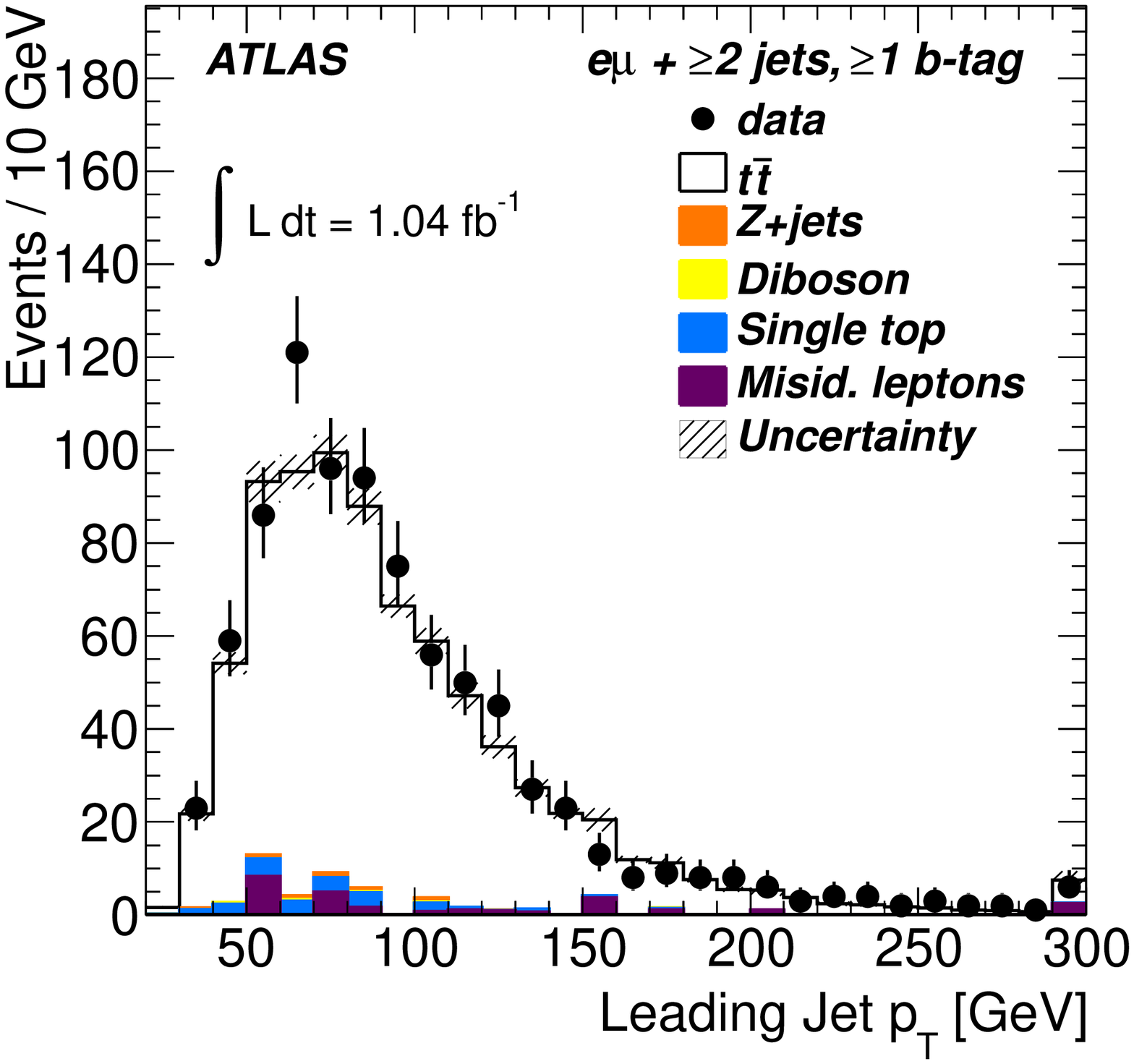}
    \end{tabular}
  \end{center}
  \caption{The top row shows (left) the $\et$ spectrum of the leading
    electron in the $ee$ channel and (right) the $\pt$ spectrum of the
    leading muon in the $\mu\mu$ channel. The bottom row shows (left)
    the missing transverse momentum in the $e\mu$ channel and (right)
    the $\pt$ spectrum of the leading jet in the $e\mu$ channel. The
    error band shows the statistical uncertainty from the Monte Carlo
    simulation. The last bin in each distribution includes the
    overflow.
    \label{fig:control_dil}}
\end{figure}

The numbers of events expected after all selection requirements are
shown in Table~\ref{tab:yield_dilepton} for the three dilepton
channels, together with the observed numbers of events in the data
sample. The observed numbers of events is in agreement with those
expected, with a purity of 94\% for all three channels. The systematic
uncertainty discussed in Section~\ref{sec:systematics} dominates the
error on the event yields, with jet energy scale, jet energy
resolution, signal and background modelling, and $b$-tagging being the
dominant sources of uncertainty. Example distributions in data are
compared with the Standard Model predictions in
Figure~\ref{fig:control_dil}. The data are described well by the
expected distributions.

\begin{table}
  \caption{Event yields in the dilepton channels after the event
    selection and event reconstruction. The table shows the expected number of
    events including their estimated total uncertainty as
    well as the number of events observed in the data sample.
    \label{tab:yield_dilepton}}
  \begin{center}    
    \begin{tabular}{llllll}
      \hline
      Process               & $ee$ channel                                   & $\mu\mu$ channel                                          & $e\mu$ channel \\
      \hline
      \ttbar\               &           159\phantom{.0} $\pm$  28            &           320\phantom{.0} $\pm$           40\phantom{.0}  &           750\phantom{.0} $\pm$          100 \\
      $Z$+jets (ee,$\mu\mu$)& \phantom{00}0.7           $\pm$ \phantom{0}2.0 & \phantom{00}2\phantom{.0} $\pm$ \phantom{0}4\phantom{.0}  & \phantom{0000}         --- \\
      $Z$+jets ($\tau\tau$) & \phantom{00}0.0           $\pm$ \phantom{0}3.8   & \phantom{00}1.4           $\pm$ \phantom{0}1.3            & \phantom{00}4.7           $\pm$ 
\phantom{00}3.1 \\
      Diboson               & \phantom{00}0.2           $\pm$ \phantom{0}0.3 & \phantom{00}0.6           $\pm$ \phantom{0}0.3            & \phantom{00}1.3           $\pm$ 
\phantom{00}0.5 \\ 
      Single top	          & \phantom{00}5.0           $\pm$ \phantom{0}1.3 & \phantom{0}10.8           $\pm$ \phantom{0}2.4            & \phantom{0}24\phantom{.0} $\pm$ 
\phantom{00}5 \\ 
      Misidentified leptons& \phantom{00}4\phantom{.0} $\pm$ \phantom{0}2   & \phantom{00}7\phantom{.0} $\pm$ \phantom{0}4\phantom{.0}  & \phantom{0}19\phantom{.0} $\pm$     
\phantom{00}9 \\
      \hline
      Total predicted	      & 170\phantom{.0} $\pm$ 30                       & 340\phantom{.0} $\pm$ 40                                  & 800\phantom{.0} $\pm$ 100 \\
      \hline
      Data                  & 191                                            & 354                                                       &  836          \\ 
      \hline
    \end{tabular}
  \end{center}
\end{table}

\section{Measurement strategies}
\label{sec:measurements}

The two methods to extract the helicity fractions are described
below. The analysis was carried out for the single-electron,
single-muon, $ee$, $e\mu$ and $\mu\mu$ channels separately. The
reconstruction of \ttbar\ events in the single-lepton channels differs
for the two methods: the template method relies more strongly on the
structure of the $\costs$ distribution than the calculation of the
angular asymmetries. It thus relies on an improved estimate of the
particle energies. A common reconstruction method was used in the
dilepton channels. The results of the analyses and their combination
are described in Section~\ref{sec:results}.

\subsection{Measurement using the template method}
\label{sec:template}

Templates for different signal and background processes were fitted to 
the observed $\costs$ distributions based on events reconstructed with a 
kinematic fit in the single-lepton 
channels~\cite{chargeasymmetry,topmass}. The fit was based on a 
likelihood which took into account the Breit-Wigner forms of the 
lineshapes of the top quark and the $W$ boson as well as the energy 
resolution of the measurements of the jets and the charged leptons. The 
missing transverse momentum was identified with the $x$- and 
$y$-components of the neutrino momentum. The top quark mass was fixed to 
172.5~\gev, and the mass of the $W$ boson was fixed to $80.4$~\gev. The 
likelihood was maximized with respect to the energies of the final state 
quarks for each association of jets to quarks. The permutations were 
weighted according to the $b$-tagging information and the weights were 
derived from the efficiency and mis-tag rate of the $b$-tagger. The 
permutation with the largest value of the likelihood was used in the 
following steps of the analysis. The overall efficiency for the 
reconstruction of the correct event topology was found to be 74\% in 
simulated \ttbar\ events. Only those events where four jets and a lepton 
are matched to partonic particles, corresponding to roughly 30\% of the 
events, were considered for the efficiency computation.

For the dilepton channels, the $t\bar{t}$ system was reconstructed by
solving a set of six independent equations:
\begin{eqnarray}
\ensuremath{p_{x}^{\nu_1} + p_{x}^{\nu_2} = \metx} ;&
\ensuremath{p_{y}^{\nu_1} + p_{y}^{\nu_2} = \mety} ;&
\ensuremath{( p_{\ell_1} + p_{\nu_1} )^{2}  = m_{W}^{2}};  \nonumber \\
\ensuremath{( p_{\ell_2} + p_{\nu_2} )^{2}  = m_{W}^{2}} ;&
\ensuremath{( p_{W_1} + p_{j_1} )^{2}     = m_{t}^{2}} ;&
\ensuremath{( p_{W_2} + p_{j_2} )^{2}      = m_{t}^{2}}.
\label{equ:a6}
\end{eqnarray}
$\metx$ and $\mety$ represent the $x$- and $y$-components of the
missing transverse momentum, $\ensuremath{p_{\ell_1}$ and
  $p_{\ell_2}}$ ($\ensuremath{p_{j_1}$ and $p_{j_2}}$, $p_{\nu_1}$ and 
$p_{\nu_2}$) correspond to the four-momenta of the two charged leptons 
(jets, neutrinos). Here, $m_{W}$ and $m_{t}$ are the $W$ boson and top 
quark masses, respectively. For events with only one $b$-tagged jet, 
this jet and the non-tagged one with the highest $\pt$ were taken as 
$b$-jets from the top quark decays. If at least two $b$-tagged jets are 
present, the two $b$-tagged jets with the largest $\pt$ in the event 
were used. The pairing of the jets with the charged leptons was based on 
the minimization of the sum of the invariant masses $m_{\ell_1 j_1}$ and 
$m_{\ell_2 j_2}$. Simulations show that this criterion gives the correct 
pairing in 68\% of the events. Up to four solutions can be found to the 
Equations~\ref{equ:a6}. The solution with the minimum product of 
neutrino transverse momenta was chosen and is motivated by the low 
$\pt$-spectrum of the neutrinos. In case no solution was found, the top 
quark mass parameter was varied in the 157.5--187.5~GeV range, to try to 
find one or more solutions. In case a range of top masses provided a 
solution, that with the top quark mass parameter closest to 172.5~GeV 
was taken. Again, multiple ambiguities were resolved by taking the 
solution with the minimum product of neutrino transverse momenta. If 
still no solution was found, the second pairing of jets and charged 
leptons was used. If no solution was found after this procedure the 
event was discarded. This happened in about 25\%{} of the events in data 
and in simulated \ttbar\ events.

\begin{figure}
  \begin{center}
    \begin{tabular}{c c}     
\includegraphics[width=0.45\textwidth]{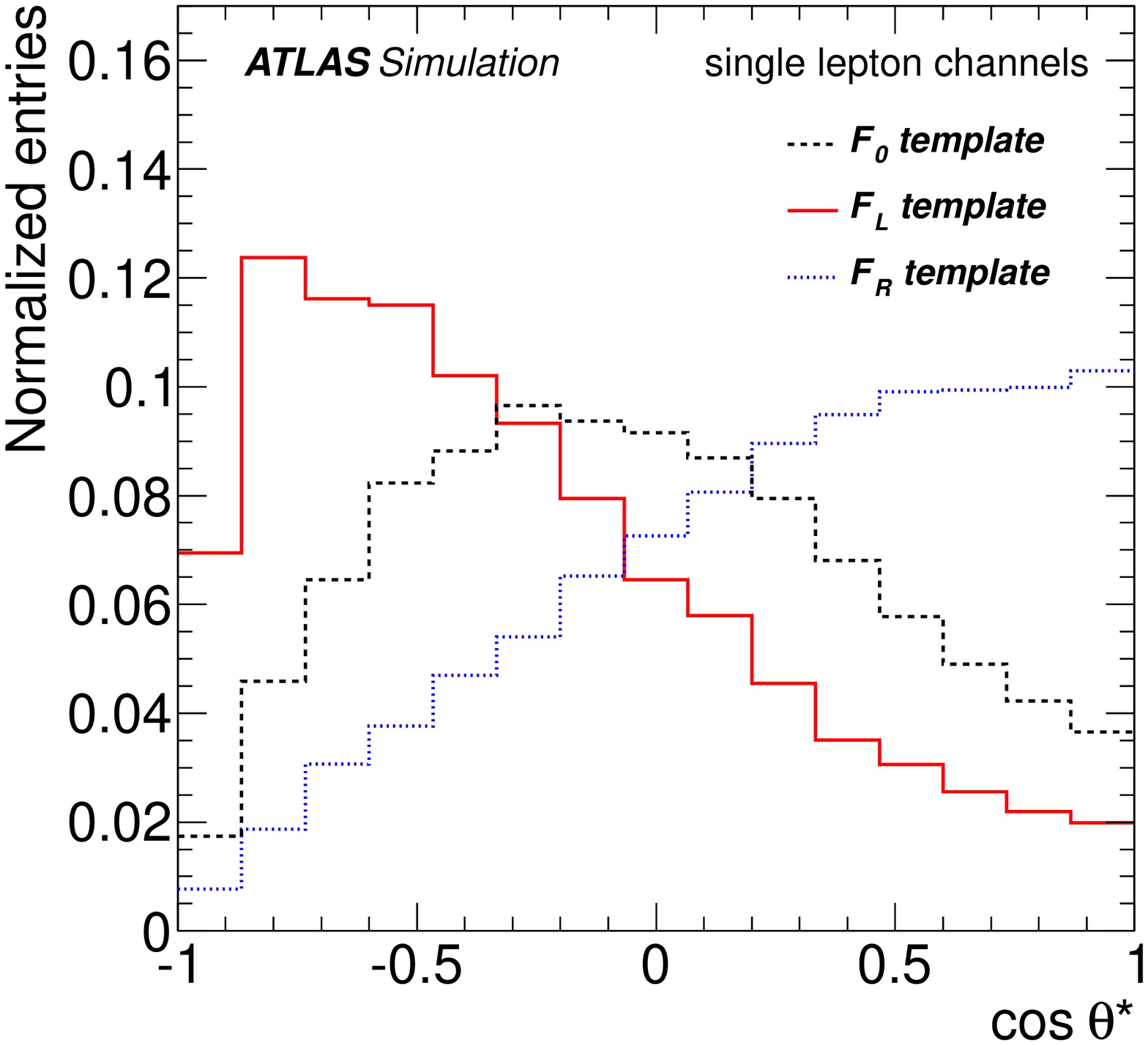} & 
\includegraphics[width=0.45\textwidth]{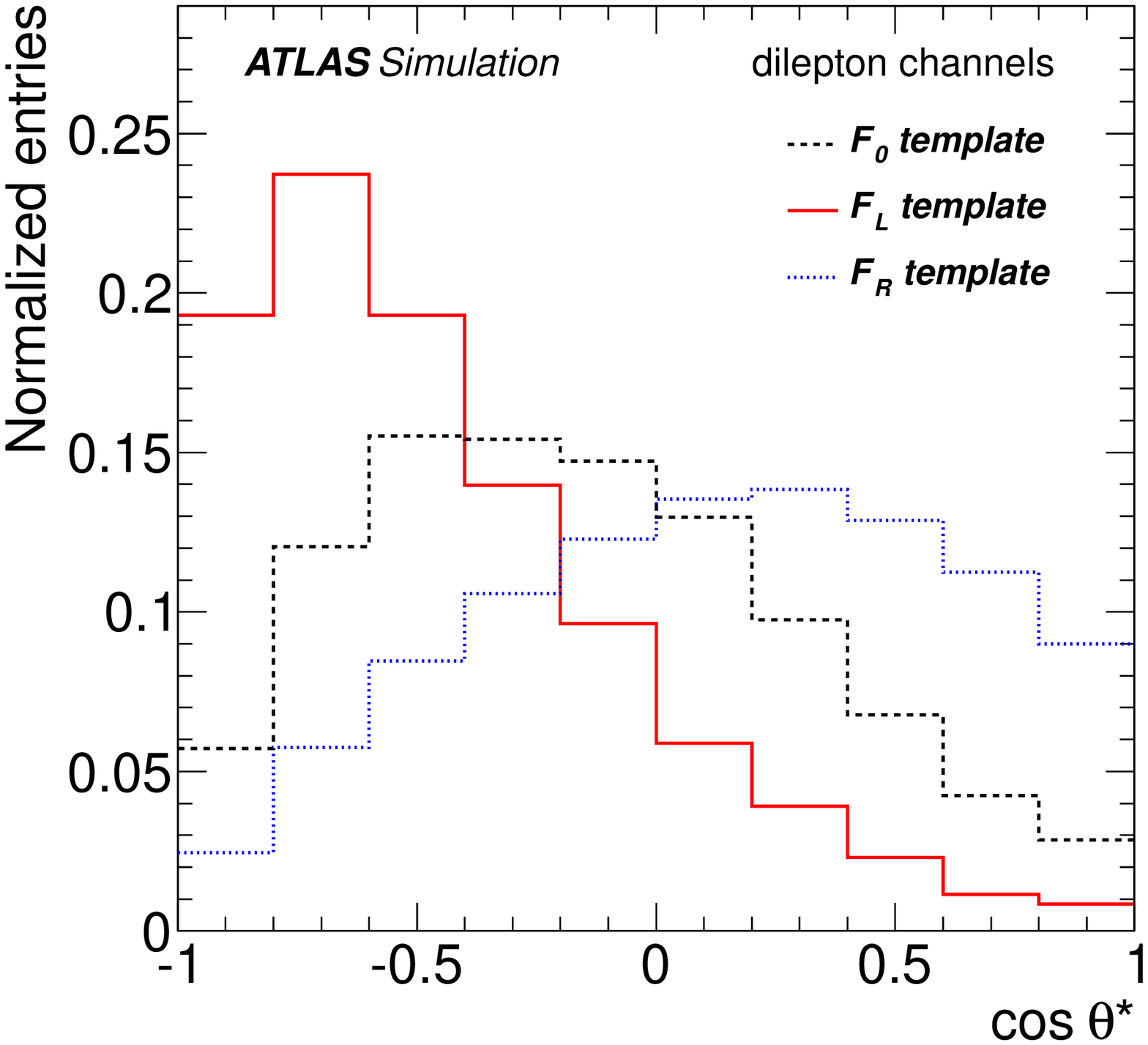} \\[-2mm]
    \end{tabular}
  \end{center}
  \caption{Distributions of $\costs$ for each of the three simulated
    signal templates.  The templates for the combined (left)
    single-lepton and (right) dilepton channels are shown.
  \label{fig:templates}}
\end{figure}

In the single-lepton channels, the following six processes were used
as templates: three different helicity state signal processes, the
background from events with misidentified leptons, the $W$+jets
contribution and the sum of all other sources of background. The three
signal templates were derived from simulation and include
single-lepton and dilepton events which pass the event
selection. These distributions are shown in
Figure~\ref{fig:templates}.  The $W$+jets template was obtained from
simulation.  The template for events with misidentified leptons was
obtained from data. Other processes contributing to the observed
spectrum include single top quark, diboson and $Z$+jets production and
were summed in a single template obtained from simulation.

In the dilepton channels, templates for the three different helicity
state signal processes were used as well as a single template
representing the different background contributions, namely single top
quark production, processes with misidentified leptons,
production of $Z$ bosons with additional jets and diboson
production. All templates were obtained from simulation, 
except the misidentified leptons template, which was obtained
from data.

A binned likelihood fit was used to estimate the expected number of
events contributing to the distribution, assuming independent
Poisson-distributed fluctuations in each bin. The number of events expected in
the $i$th bin, $\lambda_{i}$, is given as:
\begin{equation}
\lambda_{i} = \sum_{h=-1,0,+1}\lambda^{h}_{i}\cdot \epsilon^{h} + \sum_{j=1}^{N_{\mathrm{bkg}}} \lambda^{\,j}_{i} \, ,
\end{equation}
where the first term describes the expected number of signal events at
particle level with defined helicity $h$. These are reduced by
reconstruction efficiencies and acceptances summarized by the factors
$\epsilon^{h}$ estimated from simulation. The second term describes
the expected number of events from background processes and
$N_{\mathrm{bkg}}$ is the number of background templates.
The contributions of the various background templates were constrained
according to the central values and uncertainties given in
Table~\ref{tab:yield_ljets} and Table~\ref{tab:yield_dilepton}. The
uncertainties were assumed to be uncorrelated.

\subsection{Measurement using the angular asymmetries}
\label{sub:asymmetries}

The angular asymmetries were measured using the $\costs$ distribution
obtained by reconstructing the events with a $\chi^2$ minimization
technique in the single-lepton channels~\cite{CSCbook}. The
$\chi^2$ was defined according to:
\begin{equation}
 \chi^2 = \frac{(m_{\ell \nu j_a}-m_{t})^2}{\sigma_{t}^2} +
\frac{(m_{j_b j_c
j_d}-m_{t})^2}{\sigma_{t}^2} + \frac{(m_{\ell \nu}
-m_{W})^2}{\sigma_{W}^2}
+
\frac{(m_{j_c j_d}-m_{W})^2}{\sigma_{W}^2} \,\, ,
\label{chi2}
\end{equation}
where $m_{t}=172.5$~GeV, $m_{W}=80.4$~GeV, $\sigma_{t} = 14$~GeV and
$\sigma_{W} = 10$~GeV are the expected top quark and $W$ boson mass
resolutions, ${\ell}$ represents the selected electron or muon,
$m_{\ell \nu}$ is the invariant mass of the electron (muon) and the
neutrino, and $j_{a,b,c,d}$ corresponds to all possible combinations
of four jets among all selected jets in the event (with $m_{\ell \nu
  j_a}$, $m_{j_b j_cj_d}$ and $m_{\ell \nu j_a}$ being the corresponding 
invariant masses). The neutrino was reconstructed using the missing 
transverse energy, with the longitudinal component of the neutrino 
momentum ($p_z^\nu$) allowed to vary.  The solution corresponding to the 
minimum $\chi^2$ value is chosen. No $b$-tagging information was used 
in resolving the ambiguities in assigning jets.

The method described in the previous subsection was used to
reconstruct dilepton $t\bar t$ events.

For the measurement of the angular asymmetries, $A_+$ and $A_-$, the 
$\costs$ distribution was divided into four non-uniform bins, which were 
used to count the number of events above and below $z=\pm(1-2^{2/3})$, 
as defined in Equation~\ref{eq:asym}. A background subtraction in the 
observed $\costs$ distribution was performed. Subsequently, the 
following steps were applied iteratively: the number of reconstructed 
events above and below $\costs=z$ were counted in data for each 
asymmetry and correction factors were evaluated by comparing the 
Standard Model expectation with the reconstructed number of simulated 
$\ttbar$ events. These factors allowed corrections to be made for event 
selection and reconstruction effects. The obtained angular asymmetries 
were then converted into $W$ boson helicity fractions and these values 
were used to re-derive the $\costs$ distribution and evaluate new 
correction factors. The procedure was repeated until the method 
converged, {\em i.e.} until the differences between the output 
observables and the input hypothesis at the previous iteration step were 
below 0.5\%{}. Closure tests were performed using Monte Carlo samples 
with different helicity fractions and no bias was observed.


\section{Systematic uncertainties}
\label{sec:systematics}

\subsection{Sources of systematic uncertainty}

Several sources of systematic uncertainty were taken into account in
the analyses presented here. These were categorized into the modelling
of the signal and background processes, and the detector modelling. The
impact of the systematic uncertainties on the final results is
summarized in Table~\ref{tabl:systematics}.

\subsubsection{Signal and background modelling}

The signal process was modelled with different Monte Carlo generators. 
Sources of systematic uncertainty considered here were the choice of 
generator and parton shower model, the choice of parton distribution 
functions, the assumed top quark mass and the choice of parameters which 
control the amount of initial and final state radiation. Predictions 
from the {\sc MC@NLO} and {\sc Powheg} generators were compared. The 
parton showering was tested by comparing two {\sc Powheg} samples 
interfaced to {\sc Herwig} and {\sc Pythia}, respectively. The amount of 
initial and final state radiation was varied by modifying parameters in 
{\sc AcerMC} interfaced to {\sc
  Pythia}. The parameters were varied in a range comparable with those
used in the Perugia Soft/Hard tune
variations~\cite{Skands:2010ak}. The impact of the choice of parton
distribution functions was studied using the reweighting procedure
described in Ref.~\cite{CSCbook}. {\sc MC@NLO} samples were generated
assuming different top quark masses and their predictions were
compared. The observed differences in the results were scaled to
variations of 0.9~\GeV\ in the top quark mass according to the
uncertainty on its Tevatron average
value~\cite{TevatronElectroweakWorkingGroup:2011wr}. The impact of
different models of colour reconnection was studied by comparing
samples simulated with {\sc AcerMC} using the Perugia 2010 tune with
and without colour reconnection~\cite{Skands:2010ak} as well as the
tune~A-Pro and ACR-Pro~\cite{Albrow:2006rt,Buckley:2009bj}.

Background processes were either modelled by simulation or were
estimated in auxiliary measurements.  The number of events with
misidentified leptons was estimated in data for each
channel~\cite{chargeasymmetry} and the uncertainty on the
normalization was estimated to be 50\% before and 100\% after the
$b$-tagging requirement.  The normalization of $W$+jets processes was
estimated from supplementary measurements using the asymmetric
production of positively and negatively charged $W$ bosons. The
uncertainty was estimated using
Berends-Giele-scaling~\cite{Berends:1990ax} which yielded 48\% for
events with four jets and increased with the jet multiplicity by 24\%
per additional jet~\cite{Collaboration:2010ey}. Systematic
uncertainties on the shapes of the $W$+jets distributions were
assigned based on samples with different simulation parameters such as
the minimum transverse momentum of the parton and the functional form
of the factorization scale in {\sc Alpgen}. Scaling factors correcting
the fraction of heavy flavour contributions in simulated $W$+jets
samples were estimated in auxiliary measurements described in
Ref.~\cite{chargeasymmetry}. The uncertainties were 76\% for
$Wb\bar{b}$+jets and $Wc\bar{c}$+jets contributions, and 35\% for
$Wc$+jets contributions. The uncertainty on the normalization of
$Z$+jets events was estimated using Berends-Giele-scaling. The
uncertainties in the normalization were 48\% for events with four jets
and increased with the jet multiplicity by 24\% per additional jet. A
systematic uncertainty in the shape was accounted for by comparing
simulated samples generated with {\sc Alpgen} and {\sc Sherpa}. The
uncertainty on the normalization of the small background contributions
from single top quark and diboson production was estimated to be about
10\% (depending on the channel) and 5\%, respectively. The former
estimate was based on the difference between the predictions from
MC@NLO and MCFM~\cite{Campbell:2000bg}, whereas the latter comes from
scale and PDF uncertainties evaluated with MCFM.

For the template method, the bin content in each template was varied 
according to a Poisson distribution to estimate the impact of the finite 
Monte Carlo sample size used. For the measurement of the angular 
asymmetries, the $t\bar t$ reconstruction parameters were varied. In the 
single-lepton channels, the top quark and $W$ boson mass resolutions 
used in the $\chi^2$ definition were changed by 25\%{}. For the dilepton 
channels, the assumed top quark mass window used in the kinematic 
equations was reduced by 50\%{}, taking the difference in the 
observables with respect to the nominal procedure as the uncertainty 
estimate. The uncertainty due to the finite Monte Carlo sample size was 
found to be negligible since fewer bins were used in the unfolding than 
for the templates. These sources of uncertainty are labelled 
``method-specific uncertainties'' in Table~\ref{tabl:systematics}.

\subsubsection{Detector modelling}

The mis-modelling of lepton trigger, reconstruction and selection
efficiencies in simulation was corrected for by scale factors derived
from auxiliary measurements of the processes $Z\rightarrow\mu\mu$ and
$Z\rightarrow e e$~\cite{ATLAS-CONF-2011-046,Aad:2011mk}. The
uncertainties were evaluated by changing the event selection of the
supplementary measurement and by testing the stability of the results
against changing LHC and ATLAS run conditions. The same processes were
used to measure the lepton momentum scale and resolution. Scale
factors and their uncertainties were derived to match the simulation
to observed distributions. Details are given in
Ref.~\cite{Collaboration:2010ey}.

The jet energy scale was derived using information from test-beam
data, LHC collision data and simulation. Its uncertainty varies
between 2.5\% and 7\% in the central $\eta$ region, depending on jet
$\pt$ and $\eta$~\cite{Aad:2011he}. This includes uncertainties in the
flavour composition of the samples and mis-measurements from close-by
jets. An additional \pt-dependent uncertainty of up to 2.5\% was
assigned to jets matched to $b$-quarks (using Monte Carlo
generator-level information) due to differences between light-quark
and gluon jets as opposed to jets containing $b$-hadrons. Additional
uncertainties of up to 5\% (8\%) in the central (forward) region were
added due to pileup. The energy resolution for jets in Monte Carlo
simulation was adjusted to that observed in data. Uncertainties on the
energy resolution of 4--45\%, decreasing with jet $\pt$, were
assigned. The reconstruction efficiency of jets was found to be in
good agreement with the predictions from simulation and uncertainties
of 1--2\% were assigned.

The uncertainties on the momenta of electrons, muons and jets were
propagated into the missing transverse momentum. A 10\% uncertainty
from pileup was added in addition.

The $b$-tagging efficiencies and mis-tag rates have been measured in
data~\cite{ATLAS-CONF-2011-102}. Jet $\pt$-dependent scale factors,
applied to simulation to match the data, have uncertainties which
range from 9\% to 16\% and 12\% to 45\%, respectively.

The uncertainty on the measured luminosity was estimated to be
3.7\%~\cite{lumi}. 

Due to a hardware failure, a small, rectangular region of the ATLAS
electromagnetic calorimeter could not be read out in a subset of the
data (0.87~fb$^{-1}$). Data and Monte Carlo events in which a jet or
an electron were close to the affected calorimeter region were
rejected. The systematic uncertainty labelled ``calorimeter readout''
in Table~\ref{tabl:systematics} was evaluated by varying the criteria
to reject these events.


\section{Results}
\label{sec:results}

\begin{table}
  \caption{Sources of systematic uncertainty and their impact on the
    measured $W$ boson helicity fractions for the combined single-lepton and
    dilepton channels. The systematic uncertainties were symmetrized
    by using the larger uncertainty.
    \label{tabl:systematics}}
     \begin{center}
      \begin{tabular}{llll}
       \hline
       Source & \multicolumn{3}{l}{Uncertainties} \\
       & $\fz$ & $\fl$ & $\fr$ \\
       \hline
       \multicolumn{4}{l}{\it Signal and background modelling} \\
       \hline
       Generator choice           & 0.012 & 0.009 & 0.004 \\
       ISR/FSR                    & 0.015 & 0.008 & 0.007 \\
       PDF                        & 0.011 & 0.006 & 0.006 \\
       Top quark mass             & 0.016 & 0.009 & 0.008 \\
       Misidentified leptons      & 0.020 & 0.013 & 0.007 \\
       $W$+jets                   & 0.016 & 0.008 & 0.008 \\
       Other backgrounds          & 0.006 & 0.003 & 0.003 \\ 
       Method-specific uncertainties                    & 0.031 & 0.016 & 0.035 \\ 
       \hline
       \multicolumn{4}{l}{\it Detector modelling} \\
       \hline
       Lepton reconstruction      & 0.013 & 0.006 & 0.007 \\
       Jet energy scale           & 0.026 & 0.014 & 0.012 \\ 
       Jet reconstruction         & 0.012 & 0.005 & 0.007 \\  
       $b$-tagging                & 0.007 & 0.003 & 0.004 \\ 
       Calorimeter readout        & 0.009 & 0.005 & 0.004 \\
       Luminosity and pileup      & 0.009 & 0.004 & 0.005 \\
       \hline \hline                                                                                   
       Total systematic uncertainty & 0.06 & 0.03 & 0.04 \\ 
       \hline
      \end{tabular}
  \end{center}
\end{table}

\begin{figure}
  \begin{center}
    \begin{tabular}{cc}     
      \includegraphics[width=0.45\textwidth]{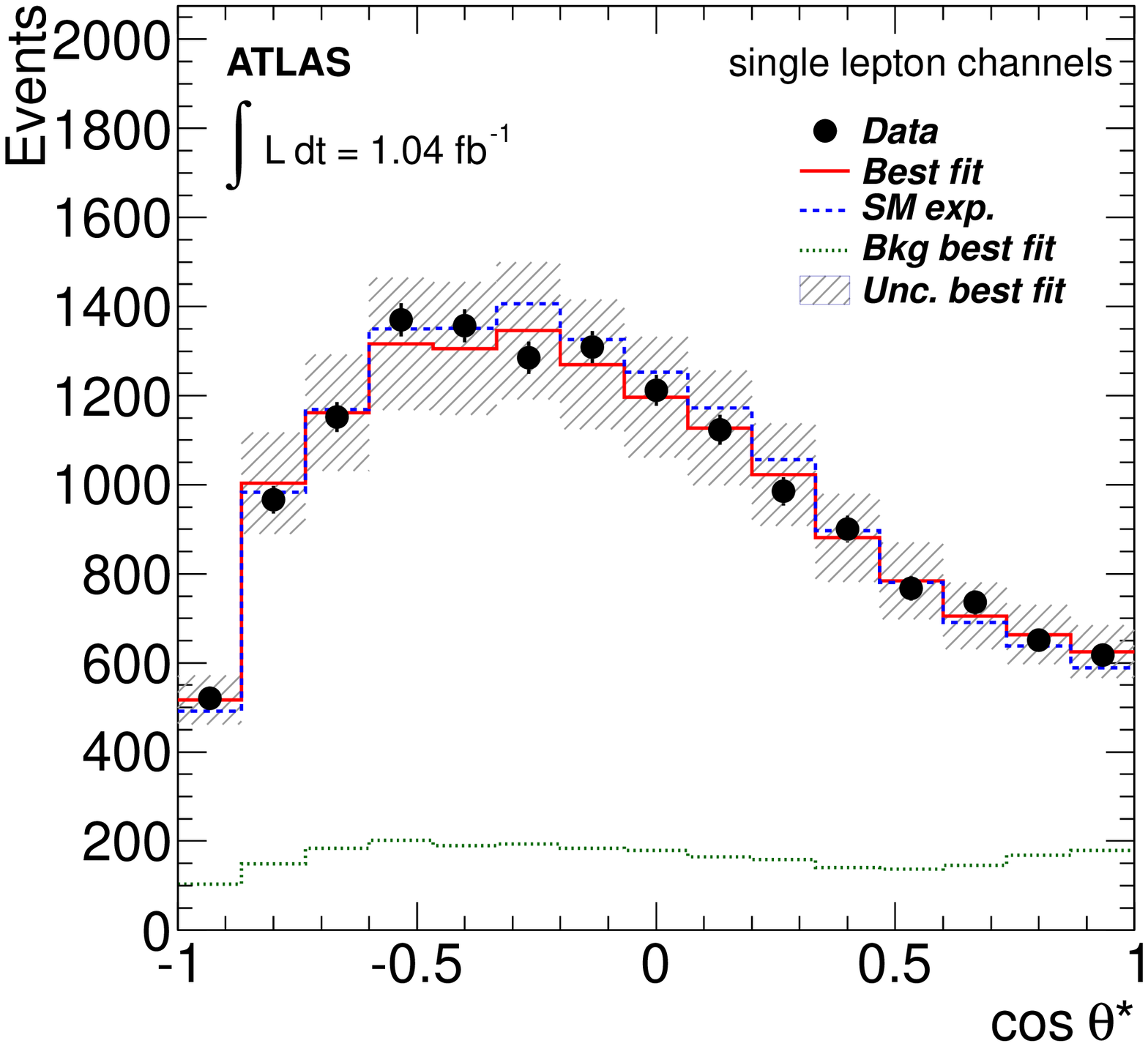} & 
      \includegraphics[width=0.45\textwidth]{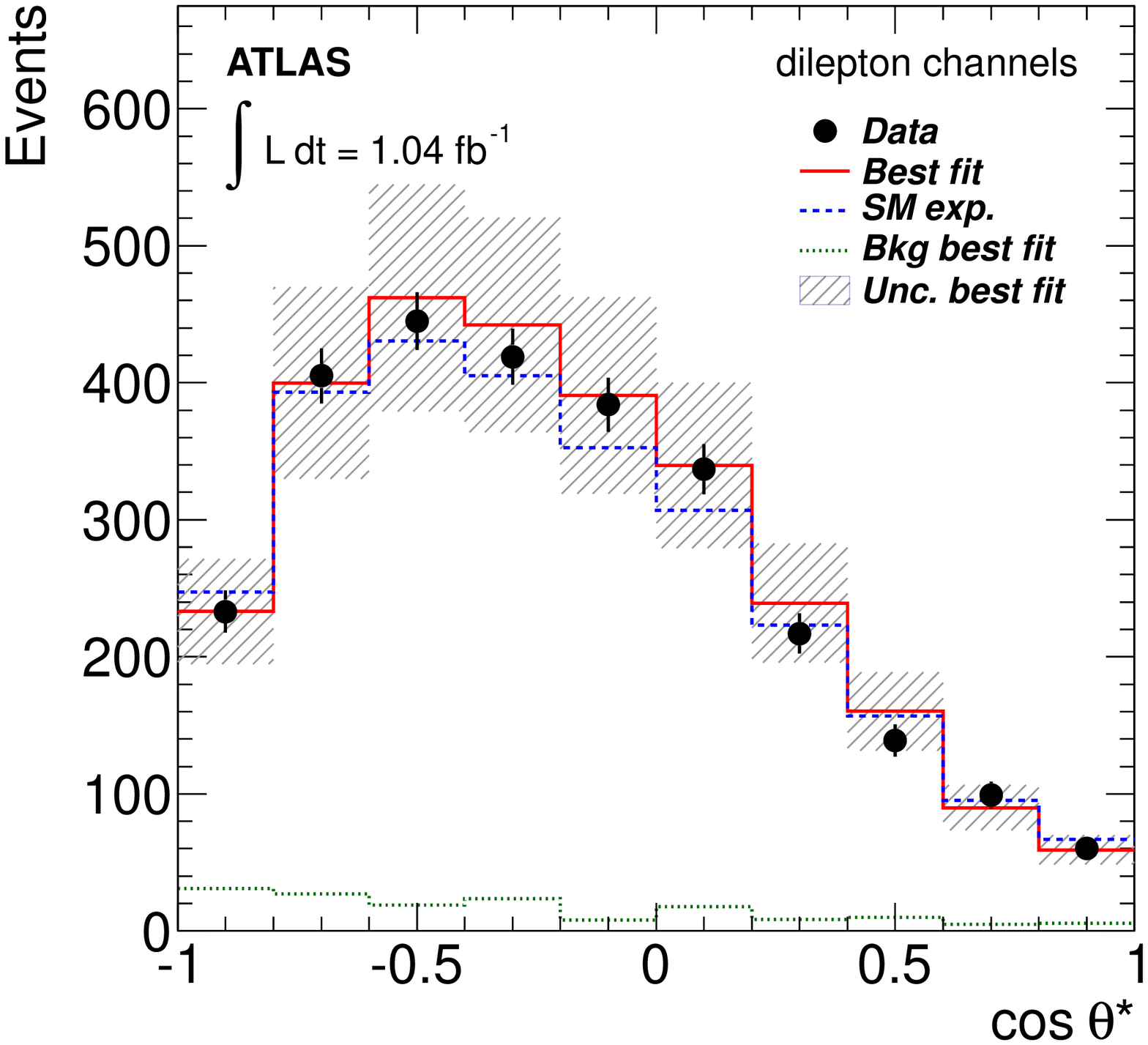} \\
    \end{tabular}
  \end{center}
  \caption{Distributions of the reconstructed $\costs$ used in the
    template method for data (markers), fitted background (dotted
    line), the Standard Model prediction (dashed line) and the best
    fit value (solid line) for the (left) single-lepton and (right)
    dilepton channels. The total uncertainties on the helicity
    fractions for the best fit values are represented by the grey
    band. For the dilepton channels, each event contributed with two
    entries, corresponding to the two leptonic decays of the $W$
    bosons.
    \label{fig:costheta_template}}
\end{figure}

\begin{figure}
  \begin{center}
    \begin{tabular}{cc}     
      \includegraphics[width=0.45\textwidth]{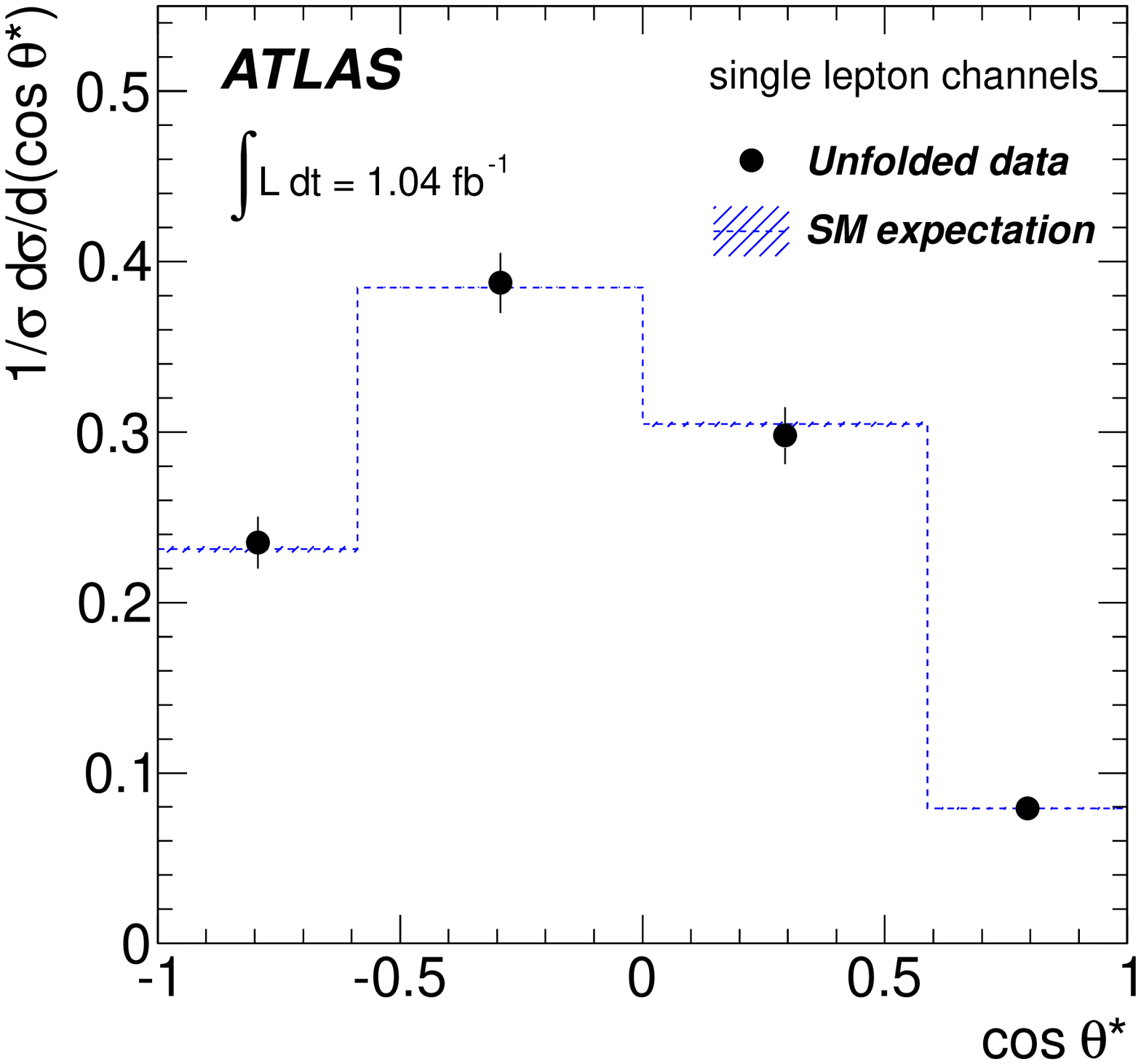} & 
      \includegraphics[width=0.45\textwidth]{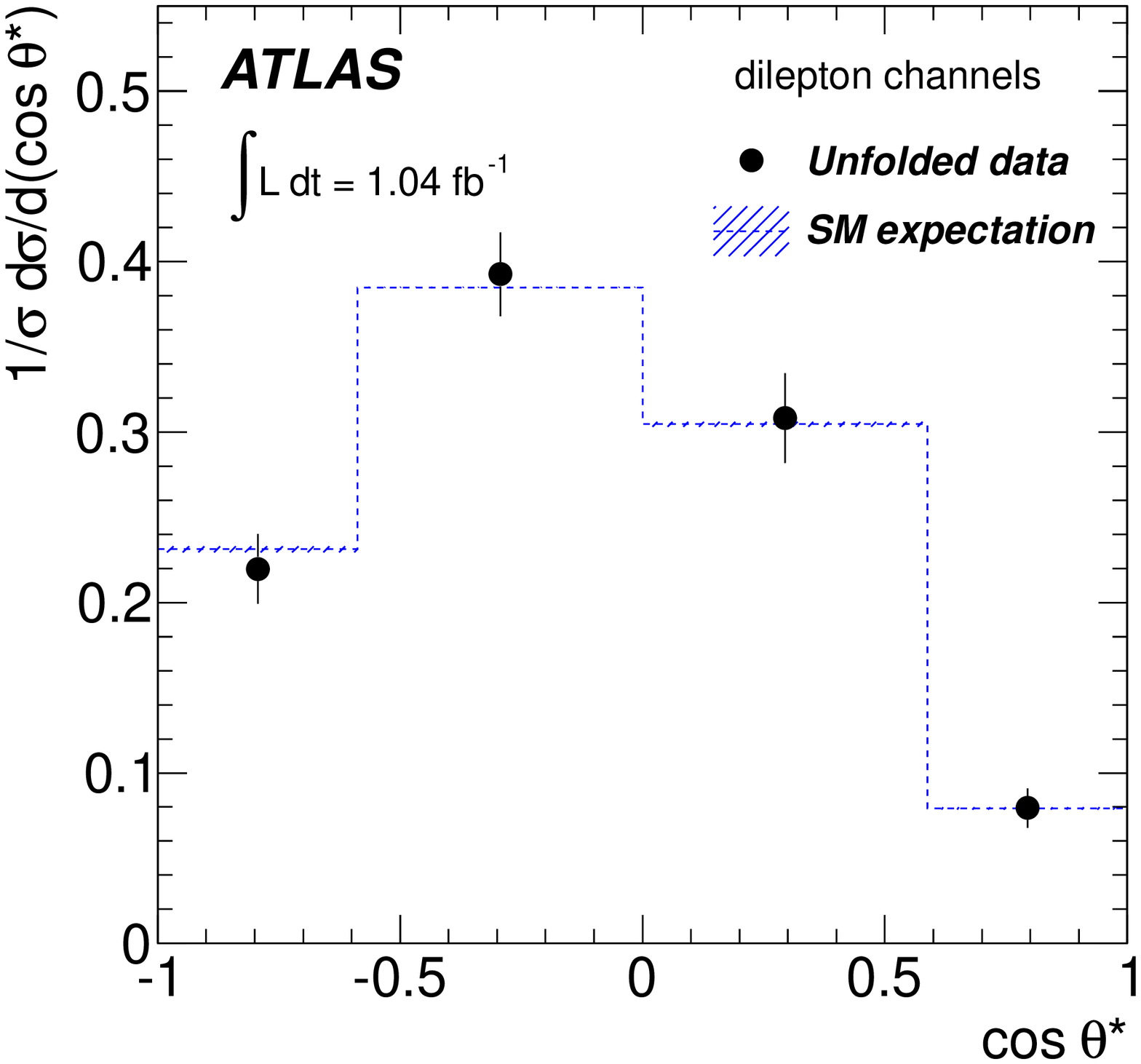} \\
    \end{tabular}
  \end{center}
  \caption{Unfolded distributions of $\costs$ for the (left)
    single-lepton and (right) dilepton channels. The error bars on
    unfolded data (markers) include both the statistical and
    systematic contributions. For comparison, the Standard Model NNLO
    QCD prediction (dashed line) and its
    uncertainty~\cite{Czarnecki:2010gb} are also shown.
    \label{fig:costheta_asymmetries}}
\end{figure}

The two methods discussed in Section~\ref{sec:measurements} were
applied to the data set described in
Section~\ref{sec:data}. Figure~\ref{fig:costheta_template} shows the
observed distribution of $\costs$ in the single-lepton and dilepton
channels together with the sum of the templates scaled to the best fit
parameters obtained from the template method as well as the Standard
Model expectations.  Figure~\ref{fig:costheta_asymmetries} shows the
distribution of $\costs$ in the single-lepton and dilepton channels
after the background subtraction and the correction for detector and
reconstruction effects as used by the asymmetry method as well as the
Standard Model expectations. 

The combination of the individual measurements of the $W$ boson
helicity fractions ($\fz$ and $\fl$) and asymmetries ($\ap$ and $\am$)
in the single-lepton and dilepton channels was done using the best
linear unbiased estimator (BLUE)
method~\cite{Lyons:1988rp,Valassi:2003mu}. The impact of the
systematic uncertainties described in Section~\ref{sec:systematics}
was studied and the results are summarized in Table~\ref{tab:results},
which also gives the combination of all channels for each method. The
results are compatible with each other and with the final
combination. In addition, the template fit was repeated with $\fr$
fixed to zero\footnote{In the evaluation of the angular asymmetries no
assumption is made for the helicity fractions, so it is not possible
to fix $\fr$ to zero.}. With the precision of the current
measurements, this differs negligibly from the Standard Model value,
and also follows the approach suggested in Ref~\cite{Zhang:2010dr}. The
results can also be found in Table~\ref{tab:results}.

The results for the angular asymmetries from the single-lepton
channels are $\ap = 0.52 \pm 0.02~\mathrm{(stat.)}  \pm
0.03~\mathrm{(syst.)}$ and $\am = -0.84 \pm 0.01~\mathrm{(stat.)}  \pm
0.02~\mathrm{(syst.)}$ whereas the results for the dilepton channels
are $\ap = 0.56 \pm 0.02~\mathrm{(stat.)} \pm 0.04~\mathrm{(syst.)}$
and $\am = -0.84 \pm 0.02~\mathrm{(stat.)}  \pm
0.04~\mathrm{(syst.)}$.

Most of the measurements and the combined result are limited by 
systematic uncertainties. The largest sources of uncertainty are the 
signal and background modelling, as well as the jet energy scale and jet 
reconstruction. The template fits are more sensitive to shape-related 
uncertainties, such as ISR/FSR and jet reconstruction, while the angular 
asymmetries are more sensitive to background normalization 
uncertainties, due to the background subtraction which needs to be 
performed.

\begin{table}
  \caption{Summary of the $W$ boson helicity fractions measured using the two
    different techniques described and the combination. The quoted
    uncertainties are the statistical (first) and the systematic (second)
    uncertainties. \label{tab:results}}
  \begin{center}
    \begin{tabular}{lccc}
      \hline
      Channel & $\fz$ & $\fl$ & $\fr$ \\
      \hline
      \multicolumn{4}{l}{\it $W$ boson helicity fractions from the template fit} \\
      \hline
      Single leptons & $0.57 \pm 0.06 \pm 0.09$ & $0.37 \pm 0.03 \pm 0.04$ & $\phantom{-}0.07 \pm 0.03 \pm 0.06$ \\
      Dileptons      & $0.92 \pm 0.10 \pm 0.10$ & $0.17 \pm 0.06 \pm 0.07$ & $-0.09           \pm 0.05 \pm 0.06$ \\
      Combination    & $0.66 \pm 0.06 \pm 0.07$ & $0.33 \pm 0.03 \pm 0.03$ & $\phantom{-}0.01 \pm 0.03 \pm 0.06$ \\ 
      \hline
      $\fr$ fixed    & $0.66 \pm 0.03 \pm 0.04$ & $0.34 \pm 0.03 \pm 0.04$ & 0 (fixed) \\
      \hline
      \multicolumn{4}{l}{\it $W$ boson helicity fractions from the angular asymmetries} \\
      \hline
      Single leptons & $0.66 \pm 0.03 \pm 0.08$ & $0.33 \pm 0.02 \pm 0.05$ & $\phantom{-}0.01 \pm 0.01 \pm 0.04$ \\
      Dileptons      & $0.74 \pm 0.06 \pm 0.10$ & $0.27 \pm 0.03 \pm 0.05$ & $-0.01           \pm 0.03 \pm 0.05$ \\
      Combination    & $0.67 \pm 0.04 \pm 0.07$ & $0.32 \pm 0.02 \pm 0.04$ & $\phantom{-}0.01 \pm 0.02 \pm 0.04$ \\ 
      \hline
      Overall combination & $0.67 \pm 0.03 \pm 0.06$ & $0.32 \pm 0.02 \pm 0.03$ & $\phantom{-}0.01 \pm 0.01 \pm 0.04$ \\ 
      \hline
    \end{tabular}
  \end{center}
\end{table}

\subsection{Combination}

\begin{figure}
  \begin{center}
    \begin{tabular}{c}
      \includegraphics[width=0.75\textwidth]{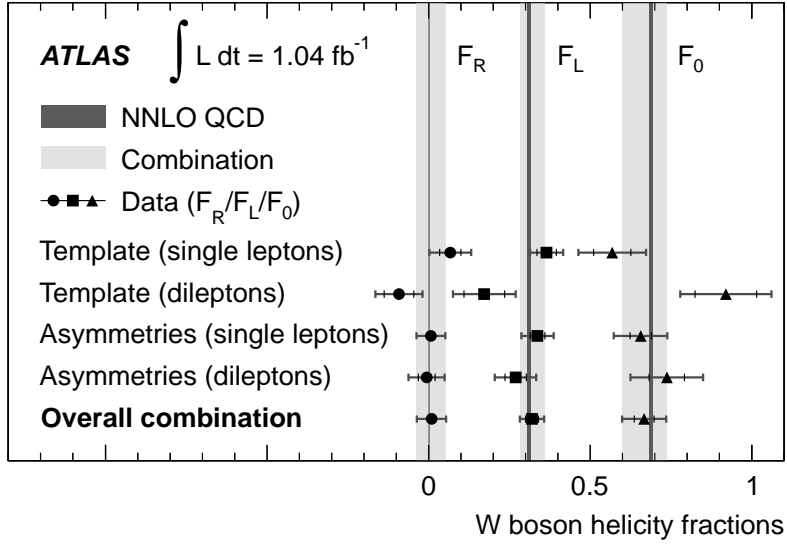}
    \end{tabular}
  \end{center}
  \caption{Overview of the four measurements of the $W$ boson helicity 
fractions and the combined values. The error bars correspond to the 
statistical and total uncertainties.
  \label{fig:combination}}
\end{figure}

The results presented in the previous sections were combined using the
BLUE method. Both the statistical correlations between analyses, and
the correlations of systematic uncertainties, were taken into
account. The statistical correlations were estimated to be between
40\%{} and 49\%{} for the single-lepton channels and between 83\%{}
and 89\%{} for the dilepton channels using pseudo-experiments obtained
from simulated samples.  The background-related systematic
uncertainties were assumed to be fully correlated within single-lepton
channels and within dilepton channels, but uncorrelated between
single-lepton and dilepton measurements. The only exception is the
uncertainty due to misidentified lepton background which depends on
the lepton flavour and thus belongs to the group of lepton-related
uncertainties, which were assumed to be fully correlated between the
channels with same flavour leptons.  The method systematic
uncertainties were assumed to be uncorrelated between channels and the
remaining sources of systematic uncertainty were assumed to be fully
correlated between channels. Various tests were performed in which the
correlations among the sources of systematic uncertainty were
varied. It was found that the values assumed for the correlations were
conservative. The systematic uncertainties on the combined values are
summarized in Table~\ref{tabl:systematics}. The four measurements of
the helicity fractions and the combined values are shown in
Table~\ref{tab:results} and Figure~\ref{fig:combination}.

The individual measurements agree reasonably well within their total
uncertainties. The $\chi^{2}/dof$ for the global combination of the
template fit and asymmetries measurements was 0.8 with a
$\chi^{2}$-probability of 75\%, where $dof$ is the number of degrees
of freedom. The largest difference between two measurements is that
between the single-lepton and dilepton channels obtained with the
template method. Since the measurements were performed in five
independent channels (single electron, single muon, $ee$, $e\mu$ and
$\mu\mu$), the combination was performed based on the five individual
measurements taking into account all correlations. The $\chi^{2}/dof$
calculated using the BLUE method for this combination was 1.3 with a
$\chi^{2}$-probability of 23\%.

The combined $W$ boson helicity fractions are:
\begin{eqnarray*}
\fz & = & 0.67  \pm 0.03~\textrm{(stat.)} \pm 0.06~\textrm{(syst.)} \, , \\
\fl & = & 0.32  \pm 0.02~\textrm{(stat.)} \pm 0.03~\textrm{(syst.)} \, , \\
\fr & = & 0.01  \pm 0.01~\textrm{(stat.)} \pm 0.04~\textrm{(syst.)} \, . 
\end{eqnarray*}
The correlation coefficient between $\fz$ and $\fl$ was estimated to be 
$-0.96$. For completeness, these results can be translated into 
angular asymmetries, yielding $\ap = 0.53 \pm 0.02$ and $\am =-0.84 \pm 
0.02$.

An alternative analysis, based on requiring two $b$-tagged jets to
further suppress the $W$+jets background and events with misidentified
leptons, was used to measure the helicity fractions in the
single-lepton channels as a cross-check. The observed $\costs$
distributions were corrected by subtracting the expected background
contributions and were unfolded using correction functions in an
iterative method similar to the one used in the measurement of the
angular asymmetries described in Section~\ref{sub:asymmetries}. The
results were found to be in agreement with the single-lepton
measurements presented in this paper.

\subsection{Constraints on the $Wtb$ vertex structure}

Any deviation of $\fz$, $\fl$, $\fr$ (or $\ap$ and $\am$) from the
Standard Model prediction could be caused by new physics contributing
to the $Wtb$ vertex.  Such new interactions associated with the top
quark may exist at higher energies. New physics can be parameterized
in terms of an effective Lagrangian~\cite{Buchmuller:1985jz} above the
electroweak symmetry breaking scale of $v=246$~\gev.  After
electroweak symmetry breaking, the
$Wtb$ Lagrangian~\cite{AguilarSaavedra:2008zc,AguilarSaavedra:2009mx}
is:
\begin{equation}
\mathcal{L}_{Wtb} = - \frac{g}{\sqrt 2} \bar b \, \gamma^{\mu} \left(
\vl P_{\rm L} + \vr P_{\rm R}
\right) t\; W_\mu^- - \frac{g}{\sqrt 2} \bar b \, \frac{i \sigma^{\mu
\nu} q_\nu}{M_W}
\left( \gl P_{\rm L} + \gr P_{\rm R} \right) t\; W_\mu^- + \mathrm{h.c.} \,,
\label{ec:lagr}
\end{equation}
where $P_{\rm L}$ ($P_{\rm R}$) is the left-handed (right-handed) 
chirality operator and
\begin{align}
& \vl = V_{tb} + C_{\phi q}^{(3,3+3)} \frac{v^2}{\Lambda^2} \,,
&&\vr = \frac{1}{2} C_{\phi \phi}^{33*} \frac{v^2}{\Lambda^2} \,,
& \gl = \sqrt 2 C_{dW}^{33*} \frac{v^2}{\Lambda^2} \,,
&& \gr = \sqrt 2 C_{uW}^{33} \frac{v^2}{\Lambda^2} \,.
\label{eq:anom_couplings}
\end{align}
The parameter $\Lambda$ is the new physics scale and $C_{\phi
  q}^{(3,3+3)}$, $C_{\phi \phi}^{33*}$, $C_{dW}^{33*}$ and
$C_{uW}^{33}$ are the effective operator
coefficients~\cite{Zhang:2010dr,AguilarSaavedra:2009mx}. The anomalous
couplings $\vr$, $\gl$, $\gr$, generated by dimension-six operators,
are absent in the Standard Model at tree level, while the coupling
$V_{tb}$ receives a correction from the operator $O_{\phi
  q}^{(3,3+3)}$.

\begin{figure}
  \begin{center}
    \begin{tabular}{c}
      \includegraphics[width=0.75\textwidth]{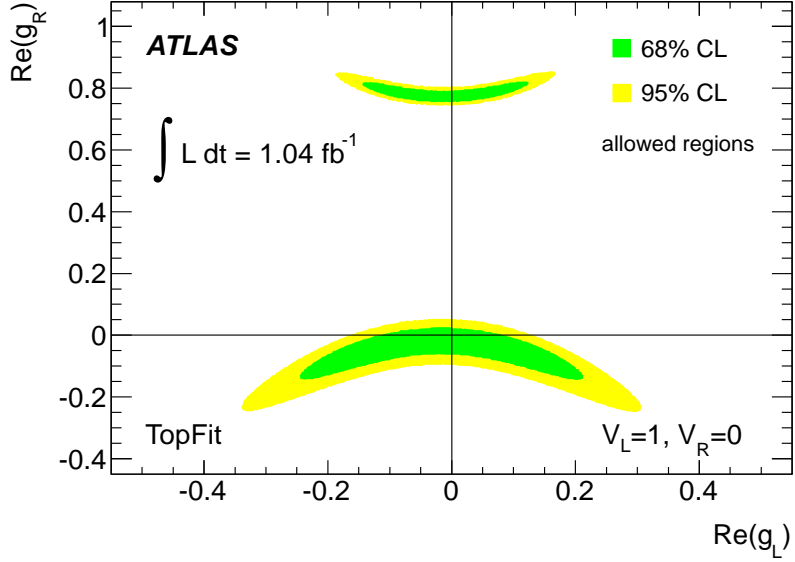}
    \end{tabular}
  \end{center}
  \caption{Allowed regions at 68\% and
    95\% confidence level (CL) for the $Wtb$ anomalous couplings 
$\gl$ 
and $\gr$. 
    In the Standard Model, the anomalous
    couplings vanish at tree level~\cite{Do:2002ky}.
  \label{fig:wtb_anom_coup}}
\end{figure}

Limits on anomalous couplings ($\vr$, $\gl$ and $\gr$) were obtained 
from the combined measurement of the $W$ boson helicity fractions by 
exploiting their dependence on these couplings, as implemented in the 
TopFit~program~\cite{AguilarSaavedra:2006fy,AguilarSaavedra:2010} and 
normalizing to $\vl = 1$. The allowed regions of $(\gl,\, \gr)$ are 
shown in Figure~\ref{fig:wtb_anom_coup}, assuming $\vr=0$.  The upper 
disconnected region in the plot shows a large-$\gr$ second solution to 
the quadratic equation relating the observables to the anomalous 
couplings. However, this region is disfavored by the measured 
cross-section for single top production at the 
Tevatron~\cite{AguilarSaavedra:2008gt,Group:2009qk,Abazov201222}.

In addition to this two-dimensional limit it is useful to set limits
on single anomalous couplings, taking only one of them non-zero at a
time. These are, at 95\% confidence level,

$$\RE (\vr) \in [-0.20,0.23] \to \frac{\RE(C_{\phi\phi}^{33})}{\Lambda^2}\in [-6.7,7.8] ~\TeV^{-2} \,,\notag$$
$$\RE (\gl) \in [-0.14,0.11] \to \frac{\RE (C_{dW}^{33})}{\Lambda^2}\in[-1.6,1.2] ~\TeV^{-2} \,,\notag$$
$$\RE (\gr) \in [-0.08,0.04] \to \frac{\RE (C_{uW}^{33})}{\Lambda^2}\in[-1.0,0.5] ~\TeV^{-2} \,. \notag$$

\noindent 
The considered $W$ boson helicity observables also allow a second region 
for $\gr$ when the remaining anomalous couplings vanish: $\RE (\gr) \in 
[0.75, 0.80]$ at 95\% confidence level. It should be noticed, however, 
that such large coupling values would imply a single top production 
cross-section value disfavored by the Tevatron 
measurements~\cite{AguilarSaavedra:2008gt,Group:2009qk,Abazov201222}. 
Using a Bayesian approach~\cite{Caldwell:2008fw}, the measurement of the 
$W$ boson helicity fractions with $\fr$ fixed at zero, was translated 
into a 95\% probability interval on $\RE (C_{uW}^{33})/\Lambda^2$, as 
proposed in~Ref.\cite{Zhang:2010dr}. This interval was found to be 
$[-0.9,2.3]~\TeV^{-2}$.

It can be seen that the limits on $C_{dW}^{33}$ (mediating the 
production of right-handed $b$-quarks in the top decay) are of the same 
order of magnitude as the limits on $C_{uW}^{33}$ (involving 
left-handed quarks). This reflects a good sensitivity to the effective 
operator corresponding to $C_{dW}^{33}$, even if its contribution is 
suppressed by $1/\Lambda^2$ instead of 
$1/\Lambda$~\cite{AguilarSaavedra:2010sq}.

These limits are more stringent than those obtained by the
D\O\ Collaboration~\cite{Abazov201221,Abazov201222}\footnote{The
  limits from the D\O\ Collaboration were derived assuming a massless
  $b$-quark.}. Indirect, model-dependent limits on the anomalous
couplings have been inferred from measurements of radiative $B$-meson
decays, measurements of $B \bar{B}$-mixing and electroweak precision
data~\cite{Grzadkowski:2008mf,delAguila:2008iz,Drobnak:2011aa,
  Drobnak:2011wj,Zhang:2012cd}, but these limits include assumptions on 
the absence of additional new physics effects that are not needed in 
this analysis.


\section{Conclusions}
\label{sec:conclusions}  

A measurement of the polarization of the $W$ bosons in top quark
decays was presented, based on 1.04~fb$^{-1}$ of data collected with
the ATLAS detector in 2011. The single-lepton and dilepton
decay topologies of top quark pairs were considered in the analysis.

The helicity fractions obtained from a combination of template fits to
the reconstructed $\costs$ distributions and angular asymmetries
calculated from the unfolded $\costs$ distributions are $\fz = 0.67
\pm 0.07$, $\fl = 0.32 \pm 0.04$ and $\fr = 0.01 \pm 0.05$. 
These results are in agreement with NNLO QCD predictions 
and are more precise than previous results obtained by the
CDF and D\O\ Collaborations~\cite{Aaltonen:2012rz,Aaltonen2010105,
Abazov:2010jn}.

Limits on the $Wtb$ vertex anomalous couplings were obtained from the 
combined results on the $W$ boson helicity fractions. These results are 
consistent with the $(V-A)$ structure of the $Wtb$ vertex and improve on 
the previously obtained limits~\cite{Abazov201222}.


\section{Acknowledgements}

We thank CERN for the very successful operation of the LHC, as well as the
support staff from our institutions without whom ATLAS could not be
operated efficiently.

We acknowledge the support of ANPCyT, Argentina; YerPhI, Armenia; ARC,
Australia; BMWF, Austria; ANAS, Azerbaijan; SSTC, Belarus; CNPq and FAPESP,
Brazil; NSERC, NRC and CFI, Canada; CERN; CONICYT, Chile; CAS, MOST and NSFC,
China; COLCIENCIAS, Colombia; MSMT CR, MPO CR and VSC CR, Czech Republic;
DNRF, DNSRC and Lundbeck Foundation, Denmark; EPLANET and ERC, European Union;
IN2P3-CNRS, CEA-DSM/IRFU, France; GNAS, Georgia; BMBF, DFG, HGF, MPG and AvH
Foundation, Germany; GSRT, Greece; ISF, MINERVA, GIF, DIP and Benoziyo Center,
Israel; INFN, Italy; MEXT and JSPS, Japan; CNRST, Morocco; FOM and NWO,
Netherlands; RCN, Norway; MNiSW, Poland; GRICES and FCT, Portugal; MERYS
(MECTS), Romania; MES of Russia and ROSATOM, Russian Federation; JINR; MSTD,
Serbia; MSSR, Slovakia; ARRS and MVZT, Slovenia; DST/NRF, South Africa;
MICINN, Spain; SRC and Wallenberg Foundation, Sweden; SER, SNSF and Cantons of
Bern and Geneva, Switzerland; NSC, Taiwan; TAEK, Turkey; STFC, the Royal
Society and Leverhulme Trust, United Kingdom; DOE and NSF, United States of
America.

The crucial computing support from all WLCG partners is acknowledged
gratefully, in particular from CERN and the ATLAS Tier-1 facilities at
TRIUMF (Canada), NDGF (Denmark, Norway, Sweden), CC-IN2P3 (France),
KIT/GridKA (Germany), INFN-CNAF (Italy), NL-T1 (Netherlands), PIC (Spain),
ASGC (Taiwan), RAL (UK) and BNL (USA) and in the Tier-2 facilities
worldwide.

\label{app:References}
\bibliographystyle{JHEP}
\bibliography{whel}

\clearpage
\begin{flushleft}
{\Large The ATLAS Collaboration}

\bigskip

G.~Aad$^{\rm 48}$,
B.~Abbott$^{\rm 111}$,
J.~Abdallah$^{\rm 11}$,
S.~Abdel~Khalek$^{\rm 115}$,
A.A.~Abdelalim$^{\rm 49}$,
O.~Abdinov$^{\rm 10}$,
B.~Abi$^{\rm 112}$,
M.~Abolins$^{\rm 88}$,
O.S.~AbouZeid$^{\rm 158}$,
H.~Abramowicz$^{\rm 153}$
% ...
\bigskip

% ...
$^{10}$ Institute of Physics, Azerbaijan Academy of Sciences, Baku, Azerbaijan\\
$^{11}$ Institut de F\'isica d'Altes Energies and Departament de F\'isica de la Universitat Aut\`onoma  de Barcelona and ICREA, Barcelona, Spain\\
% ...
$^{48}$ Fakult\"{a}t f\"{u}r Mathematik und Physik, Albert-Ludwigs-Universit\"{a}t, Freiburg i.Br., Germany\\
$^{49}$ Section de Physique, Universit\'e de Gen\`eve, Geneva, Switzerland\\
% ...
$^{88}$ Department of Physics and Astronomy, Michigan State University, East Lansing MI, United States of America\\
% ...
$^{111}$ Homer L. Dodge Department of Physics and Astronomy, University of Oklahoma, Norman OK, United States of America\\
$^{112}$ Department of Physics, Oklahoma State University, Stillwater OK, United States of America\\
% ...
$^{115}$ LAL, Universit\'e Paris-Sud and CNRS/IN2P3, Orsay, France\\
% ...
$^{153}$ Raymond and Beverly Sackler School of Physics and Astronomy, Tel Aviv University, Tel Aviv, Israel\\
% ...
$^{158}$ Department of Physics, University of Toronto, Toronto ON, Canada\\
% ...
\end{flushleft}


\end{document}